\documentclass[submission,copyright,creativecommons]{eptcs}
\providecommand{\event}{MCU 2013} %
\usepackage{amsmath}
\usepackage{amsfonts}
\usepackage{amssymb}
\usepackage{amsthm}
\usepackage{ifpdf}
\usepackage{subfig}
\usepackage{wrapfig}
\usepackage{cite}
\usepackage{comment}
\usepackage{appendix}
\usepackage[nofillcomment,noend]{algorithm2e}

\ifpdf

  \usepackage[pdftex]{epsfig}

\else

    \usepackage[dvips]{epsfig}
    \newcommand{\href}[2]{#2}

\fi

\newif\ifabstract
\newif\iffull

\ifabstract
	\fullfalse
\else
	\fulltrue
\fi

\newtoks\magicAppendix
\magicAppendix={}
\newtoks\magictoks
\newif\iflater
\laterfalse
\ifabstract
\long\def\later#1{\magictoks={#1}%
  \edef\magictodo{\noexpand\magicAppendix={\the\magicAppendix \par
    \the\magictoks%
  }}
  \magictodo}
\long\def\both#1{\magictoks={#1}%
  \edef\magictodo{\noexpand\magicAppendix={\the\magicAppendix \par
    \noexpand\setcounter{theorem-preserve}{\noexpand\arabic{theorem}}%
    \noexpand\setcounter{theorem}{\arabic{theorem}}%
    \noexpand\setcounter{section-preserve}{\noexpand\arabic{section}}%
    \noexpand\setcounter{section}{\arabic{section}}%
	\noexpand\let\noexpand\oldsection=\noexpand\thesection
	\noexpand\def\noexpand\thesection{\thesection}
	\noexpand\let\noexpand\oldlabel=\noexpand\label
	\noexpand\let\noexpand\label=\noexpand\blank
    \the\magictoks%
    \noexpand\setcounter{theorem}{\noexpand\arabic{theorem-preserve}}%
    \noexpand\setcounter{section}{\noexpand\arabic{section-preserve}}%
	\noexpand\let\noexpand\thesection=\noexpand\oldsection
	\noexpand\let\noexpand\label=\noexpand\oldlabel
  }}
  \magictodo
  \the\magictoks}
\else
\long\def\later#1{#1}
\long\def\both#1{#1}
\fi
\long\def\magicappendix{
	\latertrue%
	\the\magicAppendix%
}

\theoremstyle{definition}
\newtheorem{theorem}{Theorem}[section]
\newtheorem{lemma}{Lemma}[section]
\newtheorem{definition}{Definition}[section]

\newcommand{\Z}{\mathbb{Z}}

\newcommand{\pfunc}[3]{#1 : #2 \dashrightarrow #3 }

\newcommand{\dom}{{\rm dom} \;}

\newcommand{\res}[1]{\textrm{res}(#1)}
\newcommand{\termasm}[1]{\mathcal{A}_{\Box}[\mathcal{#1}]}
\newcommand{\prodasm}[1]{\mathcal{A}[\mathcal{#1}]}

\begin{document}

\title{On the Equivalence of Cellular Automata and the Tile Assembly Model}

\author{Jacob Hendricks \qquad\qquad Matthew J. Patitz
    \email{\hspace{-8pt}jhendric@uark.edu \qquad\qquad\hspace{10pt} patitz@uark.edu}
    \institute{Department of Computer Science and Computer Engineering, University of Arkansas}
    \thanks{This research was supported in part by National Science Foundation Grant CCF-1117672.}
}
\def\titlerunning{On the Equivalence of CA and the TAM}
\def\authorrunning{J. Hendricks \& M.J. Patitz}
\date{}
\maketitle

\begin{abstract}
In this paper, we explore relationships between two models of systems which are governed by only the local interactions of large collections of simple components:  cellular automata (CA) and the abstract Tile Assembly Model (aTAM).  While sharing several similarities, the models have fundamental differences, most notably the dynamic nature of CA (in which every cell location is allowed to change state an infinite number of times) versus the static nature of the aTAM (in which tiles are static components that can never change or be removed once they attach to a growing assembly).  We work with 2-dimensional systems in both models, and for our results we first define what it means for CA systems to simulate aTAM systems, and then for aTAM systems to simulate CA systems.  We use notions of simulate which are similar to those used in the study of intrinsic universality since they are in some sense strict, but also intuitively natural notions of simulation.  We then demonstrate a particular nondeterministic CA which can be configured so that it can simulate any arbitrary aTAM system, and finally an aTAM tile set which can be configured so that it can be used to simulate any arbitrary nondeterministic CA system which begins with a finite initial configuration.
\end{abstract}

\section{Introduction}\label{sec:intro}
Mathematical models of systems composed of large, distributed collections of simple components which are guided by only local interactions between neighboring elements have demonstrated the rise of emergent complexity, and provided enormous insight into the behaviors of many naturally occurring systems, while also guiding the modeling and development of complex artificial systems.  Two such notable models are cellular automata (CA) and the astract Tile Assembly Model (aTAM).  In this paper, we seek to explore the relationship between these two models.

Introduced by Stanislaw Ulam and John von Neumann in the 1940's, CA consist of an infinite grid of cells which can each sense their immediate neighborhoods and then all independently but synchronously update their states based on a finite set of rules and the state of their neighborhoods.  Since their introduction, CA have provided a rich theoretical framework for studying the power of systems governed by local interactions.  Much of that study has included classifications of the relative powers of various CA systems with differing neighborhoods and rules governing their state changes, and a large amount of this classification has been the result of various demonstrations of the ability of one system to simulate another, including very importantly the definitions of what it means for one system to simulate another.  A key notion developed during this study was that of intrinsic universality \cite{AlbertCulik87,DurandRoka89,DelormeMOT11,DelormeMOT11a,Ollinger-CSP08,Ollinger-STACS03,Goles-etal-2011,ArrigSchabThey}, %
which was designed to capture a strong notion of simulation, in which one particular automaton is capable of simulating the \emph{behavior} of any automaton within a class of automata. Furthermore, to simulate the behavior of another automaton, the simulating automaton must evolve in such a way that a translated rescaling (rescaled not only with respect to rectangular blocks of cells, but also with respect to time) of the simulator can be mapped to a configuration of the simulated automaton. The specific rescaling depends on the simulated automaton and gives rise to a global rule such that each step of the simulated automaton's evolution is mirrored by the simulating automaton, and vice versa via the inverse of the rule.  In this way, it is said that the simulator captures the dynamics of the simulated system, acting exactly like it, modulo rescaling.  This is in contrast to a computational simulation, for example when a general purpose digital computer runs a program to simulate a cellular automata while the processor's components don't actually arrange themselves as, and behave like, a grid of cellular automata.  Such computational simulations of computable systems can be performed by systems in any Turing universal model.  However, as we will discuss shortly, Turning universality does not imply the simulation capabilities necessary for intrinsic universality.

Introduced by Erik Winfree in 1998 \cite{Winf98}, the abstract Tile Assembly Model (aTAM) is a mathematical model in which the individual components are square ``tiles'', with ``glues'' on their edges, which are able to autonomously bind together to form structures based only on the amount and strengths of matching glues on edges of adjacent tiles.  The aTAM was inspired by Wang tiling \cite{Wang61}, but provides a model for the dynamic growth of tilings.  Like various CA, the aTAM has been proven to be computationally universal and capable of quite powerful behavior.  Recently, taking example from the work on CA, much work has been done to classify the power of the aTAM and derivative tile assembly models based on their powers of simulation \cite{USA,Versus,GeoTiles,OneTile,2HAMIU}.  In fact, \cite{IUSA} showed that the aTAM is intrinsically universal, which means that there is a single tile set $U$ such that, for any aTAM tile assembly system $\mathcal{T}$ (of any temperature), the tiles of $U$ can be arranged into a seed structure dependent upon $\mathcal{T}$ so that the resulting system (at temperature $2$), using only the tiles from $U$, will faithfully simulate the behaviors of $\mathcal{T}$.  In contrast, in \cite{2HAMIU} it was shown that no such tile set exists for the 2HAM since, for every temperature, there is a 2HAM system which cannot be simulated by any system operating at a lower temperature.  Thus no tile set is sufficient to simulate 2HAM systems of arbitrary temperature, despite the fact that the 2HAM is computationally universal, and can also simulate any arbitrary aTAM system as shown in \cite{Versus}.  Furthermore, it was shown in \cite{IUNeedsCoop} that although the aTAM in 3 dimensions is computationally universal at temperature $1$ (see \cite{CooFuSch11}), it is unable to simulate the behavior of the majority of temperature $2$ aTAM systems.  These results from \cite{IUNeedsCoop} and \cite{CooFuSch11} prove that Turing universality does not imply the simulation power necessary for intrinsic universality.

As early as the aTAM's initial introduction, its power to simulate CA was explored.  Winfree et al. showed that the 2-D aTAM can be used to simulate 1-D CA \cite{Winfree96}, and Winfree \cite{Winf98} showed that the 3-D aTAM can simulate 2-D CA.  Furthermore, the aTAM is commonly colloquially referred to as an asynchronous, nondeterministic CA in which quiescent states that change to ``tile'' states never change again (analogous to write-once memory).  These comparisons led naturally to our exploration of simulations between the two models using the same dimensions for each, namely 2-D.  However, even between systems within the same model, defining a satisfactory notion of simulation, namely one which captures the essence of one system ``behaving'' like the other while also generating analogous results, or output, can be difficult.  %
While the definition of a CA system simulating an aTAM system may be in some sense rather natural, the definition of an aTAM system simulating a CA system must take into account the write-once nature of tile assembly systems.  To account for this, we modify the standard notions of simulation used in intrinsic universality to allow for an increasing scale factor during simulation.  Essentially, such simulation definitions typically make use of a standard block replacement scheme in which, throughout the simulation, each constant sized block of the simulator can be directly mapped to an element of the simulated system.  To allow a static model such as the aTAM to simulate a dynamic model such as CA, we allow the scale factor of the simulation to increase after each time step of the simulated system is completed.

For our main results, we present the following.  First, a single nondeterministic, synchronous CA which, for any arbitrary aTAM system $\mathcal{T}$, can be given an initial configuration dependent upon $\mathcal{T}$ so that it will exactly simulate $\mathcal{T}$, producing the same output patterns (modulo rescaling) and preserving the dynamics of $\mathcal{T}$.  Second, we exhibit a single aTAM tile set which, for any nondeterministic, synchronous CA $\mathcal{C}$ which begins with a finite initial configuration (i.e. all but a finite number of cells begin in a quiescent state), can be given an initial seed configuration dependent upon $\mathcal{C}$ so that it will exactly simulate $\mathcal{C}$, producing the same output patterns (modulo rescaling) and preserving the dynamics of $\mathcal{C}$.

\section{Preliminaries}
Here we define the terms and models used throughout the rest of the paper.

We work in the $2$-dimensional discrete space $\Z^2$. Define the set
$U_2 = \{(0,1), (1,0), (0,-1), (-1,0)\}$ to be the set of all
\emph{unit vectors} in $\mathbb{Z}^2$.
We also sometimes refer to these vectors by their
cardinal directions $N$, $E$, $S$, $W$, respectively.
All \emph{graphs} in this paper are undirected.
A \emph{grid graph} is a graph $G =
(V,E)$ in which $V \subseteq \Z^2$ and every edge
$\{\vec{a},\vec{b}\} \in E$ has the property that $\vec{a} - \vec{b} \in U_2$.

In the subsequent definitions, given two partial functions $f,g$, we write $f(x) = g(x)$ if~$f$ and~$g$ are both defined and equal on~$x$, or if~$f$ and~$g$ are both undefined on $x$.

\subsection{The abstract Tile Assembly Model}
\label{sec:atam-def}
In this section we give an informal description of the abstract Tile Assembly Model (aTAM), which is the theoretical version of the TAM which does not model the kinetics of physical self-assembling systems. The reader is encouraged to see \cite{RotWin00, Winf98, jSSADST} for a formal development of the model.

Intuitively, a tile type $t$ is a unit square that can be
translated, but not rotated, having a well-defined ``side
$\vec{u}$'' for each $\vec{u} \in U_2$. Each side $\vec{u}$ of $t$
has a ``glue'' with ``label'' $\textmd{label}_t(\vec{u})$--a string
over some fixed alphabet--and ``strength''
$\textmd{str}_t(\vec{u})$--a nonnegative integer--specified by its type
$t$. Two tiles $t$ and $t'$ that are placed at the points $\vec{a}$
and $\vec{a}+\vec{u}$ respectively, \emph{bind} with \emph{strength}
$\textmd{str}_t\left(\vec{u}\right)$ if and only if
$\left(\textmd{label}_t\left(\vec{u}\right),\textmd{str}_t\left(\vec{u}\right)\right)
=
\left(\textmd{label}_{t'}\left(-\vec{u}\right),\textmd{str}_{t'}\left(-\vec{u}\right)\right)$.
Here the glue function is assumed to be the usual diagonal glue function. In other words,
only glues with matching labels are allowed to interact.

Fix a finite set $T$ of tile types.
A $T$-\emph{assembly}, sometimes denoted simply as an \emph{assembly} when $T$ is clear from the context, is a partial
function $\pfunc{\alpha}{\Z^2}{T}$ defined on at least one input, with points $\vec{x}\in\Z^2$ at
which $\alpha(\vec{x})$ is undefined interpreted to be empty space,
so that $\dom \alpha$ is the set of points with tiles.
We write $|\alpha|$ to denote $|\dom \alpha|$, and we say $\alpha$ is
\emph{finite} if $|\alpha|$ is finite. For assemblies $\alpha$
and $\alpha'$, we say that $\alpha$ is a \emph{subassembly} of
$\alpha'$, and write $\alpha \sqsubseteq \alpha'$, if $\dom \alpha
\subseteq \dom \alpha'$ and $\alpha(\vec{x}) = \alpha'(\vec{x})$ for
all $x \in \dom \alpha$.

For $\tau \in \mathbb{N}$, an assembly is \emph{$\tau$-stable} if every cut of its binding graph has strength at least $\tau$, where the weight of an edge is the strength of the glue it represents.
That is, the assembly is stable if at least energy $\tau$ is required to separate the assembly into two parts.
In the aTAM, self-assembly begins with a \emph{seed assembly} $\sigma$ (typically
assumed to be finite and $\tau$-stable) and
proceeds asynchronously and nondeterministically, with tiles
adsorbing one at a time to the existing assembly in any manner that
preserves stability at all times.

An aTAM \emph{tile assembly system} (\emph{TAS}) is an ordered triple
$\mathcal{T} = (T, \sigma, \tau)$, where $T$ is a finite set of tile
types, $\sigma$ is a seed assembly with finite domain, and $\tau$ is
the temperature. An \emph{assembly sequence} in a TAS $\mathcal{T} = (T, \sigma, \tau)$ is
a (possibly infinite) sequence $\vec{\alpha} = ( \alpha_i \mid 0
\leq i < k )$ of assemblies in which $\alpha_0 = \sigma$ and each
$\alpha_{i+1}$ is obtained from $\alpha_i$ by the ``$\tau$-stable''
addition of a single tile. The \emph{result} of an assembly sequence
$\vec{\alpha}$ is the unique assembly $\res{\vec{\alpha}}$
satisfying $\dom{\res{\vec{\alpha}}} = \bigcup_{0 \leq i <
k}{\dom{\alpha_i}}$ and, for each $0 \leq i < k$, $\alpha_i
\sqsubseteq \res{\vec{\alpha}}$.

We write $\prodasm{T}$ for the
\emph{set of all producible assemblies of} $\mathcal{T}$. An
assembly $\alpha$ is \emph{terminal}, and we write $\alpha \in
\termasm{\mathcal{T}}$, if no tile can be stably added to it. We
write $\termasm{T}$ for the \emph{set of all terminal assemblies of
} $\mathcal{T}$. 
The set $\prodasm{T}$ is partially ordered by the relation $\longrightarrow$ defined by
\begin{eqnarray*}
\alpha \longrightarrow \alpha' & \textmd{iff} & \textmd{there is an assembly sequence } \vec{\alpha} = (\alpha_0,\alpha_1,\ldots) \\
                               &              & \textmd{such that } \alpha_0 = \alpha \textmd{ and } \alpha' = \res{\vec{\alpha}}. 
\end{eqnarray*}
A TAS ${\mathcal T}$ is \emph{directed}, or
\emph{produces a unique assembly}, if it has exactly one terminal
assembly i.e., $|\termasm{T}| = 1$. The reader is cautioned that the
term ``directed'' has also been used for a different, more specialized
notion in self-assembly \cite{AKKRS09}. We interpret ``directed'' to
mean ``deterministic'', though there are multiple senses in which a
TAS may be deterministic or nondeterministic.

\subsection{Cellular Automata}
In our discussion about cellular automata, we will use the following definitions. Most of the conventions used in these definitions come from \cite{DelormeMOT11}.

\begin{definition}\label{CADef}
A {\em $2$-dimensional nondeterministic cellular automata} $\mathcal{A}$ is a $4$-tuple $(\mathbb{Z}^2, S, N, \delta)$ where
\begin{enumerate}
  \item $S$ is a finite set of states.
  \item $N\subset \mathbb{Z}^2$ is a finite set defining the neighborhood of $\mathcal{A}$.
  \item $\delta : S^{|N|} \to 2^S$ is the local rule of $\mathcal{A}$. $\delta$ maps a neighborhood defined by $N$ and a point in $\mathbb{Z}^2$, usually referred to as a {\em cell}, to a set of states.
\end{enumerate}

Note that a {\em deterministic cellular automata} is simply a special case of a nondeterministic CA in which $\delta: S^{|N|} \to S$, i.e. it maps each neighborhood to a single state.

A {\em configuration} $c$ is a mapping from $\mathbb{Z}^2$ to $S$. Let $C$ be the set of configurations in $\mathcal{A}$.
The {\em global rule} $G$ is obtained as follows.
For $p\in \mathbb{Z}^2$, $G:C\to 2^C$ such that $c^{\prime} \in G(c) \iff c^{\prime}(p) \in \delta(c_{p+v_1}, \dots, c_{p+v_k})$ where $\{v_1, \dots, v_k\} = N$.
\end{definition}

We assume that $S$ contains a unique {\em quiescent state} where a quiescent state $q$ is a state such that $\delta$ maps a neighborhood of cells in this state to a singleton set containing only the quiescent state. In this paper, we only consider finite initial configurations (which we will typically denote by $c_0$) where all but finitely many cells are quiescent. In this paper we are concerned with the CA-initial configuration pair $\left(\mathcal{A}, c_0\right)$ and refer to such pairs as CA {\em systems}.

There are many interesting examples of cellular automata. One of particular interest here is John Conway's Game of Life. (See \cite{Gardner70}.)
This is a $2$D cellular automata where each cell is in one of two states $\mathtt{alive}$ or $\mathtt{dead}$.
Local rules are given for a $3\times 3$ squares for cells according to the following.
\begin{enumerate}
 \item  An $\mathtt{alive}$ cell with less than two neighbors becomes $\mathtt{dead}$.
 \item  An $\mathtt{alive}$ cell with two or three neighbors stays  $\mathtt{alive}$.
 \item  An $\mathtt{alive}$ cell with more than three neighbors becomes $\mathtt{dead}$.
 \item  A $\mathtt{dead}$ cell with three $\mathtt{alive}$ neighbors becomes $\mathtt{alive}$.
\end{enumerate}
These simple rules give rise to an amazing amount of complexity and structure. In fact, in~\cite{Rendell11} a universal Turing machine built in Conway's Game of Life is presented that starts from a finite configuration that encodes another Turing machine and its tape and simulates the execution of the encoded Turing machine with the encoded tape as input.

\subsection{CA simulation of a TAS}
For $S$ as in Definition~\ref{CADef} and $k$ a vector of $\mathbb{Z}^2$, let $\psi^k : S^{\mathbb{Z}^2}\to S^{\mathbb{Z}^2}$ be the bijection mapping a configuration $c$ to the configuration
$c^{\prime}$ such that for each cell $i$, $c^{\prime}_{i+k} = c_i$. $\psi^k$ is called the {\em shift} operator.
Now let $m=(m_1, m_2)$ be a pair of strictly positive integers.
$o^m: S^{\mathbb{Z}^2}\to (S^{[0,m_1]\times[0,m_2]})^{\mathbb{Z}^2}$ is the bijection such that for all $c\in S^{\mathbb{Z}^2}$, $z\in \mathbb{Z}^2$ and $p\in [0,m_1]\times[0,m_2]$,
$o^m(c)(z)(p) = c(mz + p)$. $o^m$ is called the {\em packing} map.
Let $\mathcal{A} = (\mathbb{Z}^2, S, N, \delta)$ be a cellular automaton. An {\em $\langle m,n,k\rangle$-rescaling} of $\mathcal{A}$ is a cellular automaton $\mathcal{A}^{\langle m,n,k \rangle}$
with states set $S^{[0,m_1]\times[0,m_2]}$ and global transition function $o^m\circ\psi^k\circ G_{\mathcal{A}}^{n} \circ o^{-m}$, where $G^n_{\mathcal{A}}$ is the composition of the global function for $\mathcal{A}$ $n$ times.

We now define what it means to say that a synchronous nondeterministic $2$D cellular automata with an initial configuration {\em simulates} an aTAM system. First we let $R$ be the partial function that maps individual cells in some state to single tiles with some tile type. $R$ is the {\em representation} function.
In the following definitions, $\mathcal{A} = (\mathbb{Z}^2, S, N, \delta)$ is a synchronous nondeterministic CA with $C$ denoting the set of configurations and $\mathcal{T} = (T, \sigma, \tau)$ is an aTAM system.
We denote by $c_0$ a finite initial configuration in $C$ and let $\tilde{C} = \cup_{n=0}^{\infty}G^n(c_0)$. In other words, $\tilde{C}$ is all of the configurations obtained by applying the global rule some number of times to $c_0$.
Let $R^*:\tilde{C} \to \prodasm{T}$ be the canonical extension of $R$. Finally, we let $\left(\mathcal{A}, c_0\right)$ be the pair consisting of the CA $\mathcal{A}$ and the initial configuration $c_0$.

\begin{definition}\label{def:follows}
We say that $\mathcal{T}$ {\em follows}  $\left(\mathcal{A}, c_0\right)$ iff
for all $c \in \tilde{C}$, $\alpha, \beta \in \mathcal{A}[\mathcal{T}]$ and $n\geq 0$, if $R^*(c) = \alpha$ and $\beta \in R^*[G^n(c)]$ then $\alpha \longrightarrow \beta$.
\end{definition}

Note that $R^*[G^n(c)]$ denotes the image of the set $G^n(c)$ under $R^*$. Informally, Definition~\ref{def:follows} means that if a configuration represents an assembly $\alpha$, then anything this configuration maps to under applications of the global rule represents some assembly that $\alpha$ can grow into.
The following definition captures the idea that for an assembly $\alpha$ represented by a configuration $c$, any assembly that $\alpha$ grows into is represented by a configuration obtained from $c$ by applications of the
global rule.

\begin{definition}
We say that  $\left(\mathcal{A}, c_0\right)$ {\em models} $\mathcal{T}$ if
$\forall \alpha \in \mathcal{A}[\mathcal{T}], \exists c \in \tilde{C}$ such that $R^*(c) = \alpha$ and $\forall \beta \in \mathcal{A}[\mathcal{T}]$,
if $\alpha \longrightarrow \beta$ then $\exists n\geq 0$ such that $\beta \in R^*[G^n(c)]$.
\end{definition}

Note that a configuration representing some terminal assembly $\alpha$ must transition to configurations that still represent $\alpha$.
Finally, we give the definition of simulation.

\begin{definition}
 $\left(\mathcal{A}, c_0\right)$ {\em simulates} $\mathcal{T}$ iff there is an ${\langle m,n,k \rangle}$-rescaling $\mathcal{A^{\prime}}$ of $\mathcal{A}$ such that $\mathcal{T}$ follows $\mathcal{A^{\prime}}$ and $\mathcal{A^{\prime}}$ models $\mathcal{T}$.
\end{definition}

\subsection{TAS simulation of a CA}

As in the previous section, $\mathcal{A} = (\mathbb{Z}^2, S, N, \delta)$ denotes a synchronous nondeterministic CA with a finite initial configuration $c_0$, $C$ denotes the set of configuration and $\mathcal{T} = (T, \sigma, \tau)$ denotes an aTAM system. Again, let $\tilde{C} = \cup_{n=0}^{\infty}G^n(c_0)$ where $c_0$ is the finite initial configuration of $\mathcal{A}$ and let $\left(\mathcal{A}, c_0\right)$ be the pair consisting of the CA $\mathcal{A}$ and the initial configuration $c_0$.

Because any aTAM system produces static assemblies and the state of a cell of a CA may change multiple times, it would be impossible to represent a cell of a configuration in $\tilde{C}$ with fixed block assemblies over $T$. Therefore, we
introduce the notion of a {\em scalable representation function}.

For $n\in\Z^+$, an \emph{$n$-block supertile} over $T$ is a partial function $\alpha : \Z_n^2 \dashrightarrow T$, where $\Z_n = \{0,1,\ldots,n-1\}$.
Let $B^T_n$ be the set of all $n$-block supertiles over $T$.  The $n$-block with no domain is said to be $\emph{empty}$.  For a general assembly $\alpha:\Z^2 \dashrightarrow T$, define $\alpha^n_{x,y}$ to be the $n$-block supertile defined by $\alpha^n_{x,y}(i,j) = \alpha(nx+i,ny+j)$ for $0 \leq i,j< n$.

Let $R_n$ for $n\in \mathbb{N}$ be a partial function that maps assemblies over $T$ to configurations in $C$ with the following property. If $\alpha \in \prodasm{T}$ and $R_n(\alpha) = c$, then for some $n$, $\alpha$ can be broken into $n$-block supertiles such that $R_n$ maps these supertiles to cells of $c$. In other words, for a given assembly $\alpha$, the partial function $R_n$ either is not defined on $\alpha$ or maps $\alpha$ to $c\in C$ by mapping $n$-block supertiles of $\alpha$ to cells of $c$.
Then the scalable representation function
is defined as $R:\mathbb{N}\times \prodasm{T} \dashrightarrow \tilde{C}$ where $R(n,\beta)=R_n(\beta)$.
Finally, we define simulation of a CA with initial configuration $c_0$ by an aTAM system.

\begin{definition}\label{def:tassim}
$\mathcal{T}$ {\em simulates} $\left(\mathcal{A}, c_0\right)$ (under scalable representation function $R$) iff
there exists a computable function $f:\mathbb{N}\to \mathbb{N}$ such that the following hold.
\begin{enumerate}
 \item $\forall n\in \mathbb{N}$, $R_{f(n)}[\prodasm{T}] = G^n(c_0)$.\label{def:statement1}
 \item $\forall \alpha\in \prodasm{T}$ such that $R_{f(n)}(\alpha) = c_n\in G^n(c_0)$, for any $\beta \in \prodasm{T}$ in the domain of $R_{f(n+1)}$ such that $\alpha \longrightarrow \beta$ it must be the case that $R_{f(n+1)}(\beta) \in G^{n+1}(c_0)$. \label{def:statement2}
 \item $\forall c_n\in G^n(c_0)$ such that $R_{f(n)}(\alpha) = c_n$ for some $\alpha \in \prodasm{T}$, if $\alpha \longrightarrow \beta$ where $\beta$ is in the domain of $R_{f(n+1)}$ then
$R_{f(n+1)}(\beta) \in G^{n+1}(c_0)$.\label{def:statement3}
\end{enumerate}

\end{definition}

In Definition~\ref{def:tassim}, $f$ can be thought of as taking a time step $n$ and determines a block size for the representation.
Then $R$ takes $f(n)$ and an assembly and either returns a configuration in $G^n(c_0)$ or nothing if the assembly has not fully simulated the $n^{th}$ time step.  This is necessary to simulate the dynamics of a synchronous CA, in which all cells simultaneously update their states.
Basically statement~\ref{def:statement1} of Definition~\ref{def:tassim} says that starting with an initial configuration, every configuration obtained by applying the global rule is represented by some assembly over $T$ and that any step-assembly pair $(n,\alpha)$ in the domain of $R$ represents some configuration.
Moreover, statements \ref{def:statement2} and \ref{def:statement3} of Definition~\ref{def:tassim} implies that these representations follow the action of the global rule.

\section{A Nondeterministic CA Which Can Simulate Any aTAM System}\label{CAIU4aTAM}

In Theorem~\ref{thm:caiu}, we show that for any aTAM system, there is a synchronous nondeterministic CA such that for an appropriate choice of finite initial configuration, this CA simulates the aTAM system. This gives some sense of a synchronous nondeterministic CA being {\em intrinsically universal for} the aTAM.

\begin{theorem}\label{thm:caiu}
There exists a synchronous nondeterministic CA $\mathcal{A} = (\mathbb{Z}^2, S, N, \delta)$ such that for any aTAM system $\mathcal{T} = (T, \sigma, \tau)$ there exists a finite initial configuration
$c_0$ of $\mathcal{A}$ so that $\left(\mathcal{A}, c_0\right)$ simulates $\mathcal{T}$.
\end{theorem}

To show this Theorem, we appeal to the following Lemma. The construction in Section~\ref{sec:caconstruction}, proves this Lemma.
\begin{lemma}\label{lem:casim}
For any aTAM system $\mathcal{T} = (T, \sigma, \tau)$, there exists a synchronous nondeterministic CA $\mathcal{A} = (\mathbb{Z}^2, S, N, \delta)$ and an initial configuration $c_0$ such that $\left(\mathcal{A}, c_0\right)$ simulates $\mathcal{T}$.
\end{lemma}

Then Theorem~\ref{thm:caiu} is proven as follows. First, there is a tile set $U$ which is intrinsically universal for the aTAM and can be used at temperature $\tau=2$ to simulate any aTAM system. Therefore we let $\mathcal{U}$ be an aTAM system that uses $U$ at $\tau=2$.
By Lemma~\ref{lem:casim},  we can
then give a CA that suffices for Theorem~\ref{thm:caiu} by constructing a CA that simulates an arbitrary $\mathcal{U}$ (i.e. one with an arbitrary seed). See Section~\ref{sec:pocCAsimTAS} for more details.

\subsection{CA Construction}\label{sec:caconstruction}
\begin{figure}[htp]
\begin{center}
\includegraphics[width=1.2in]{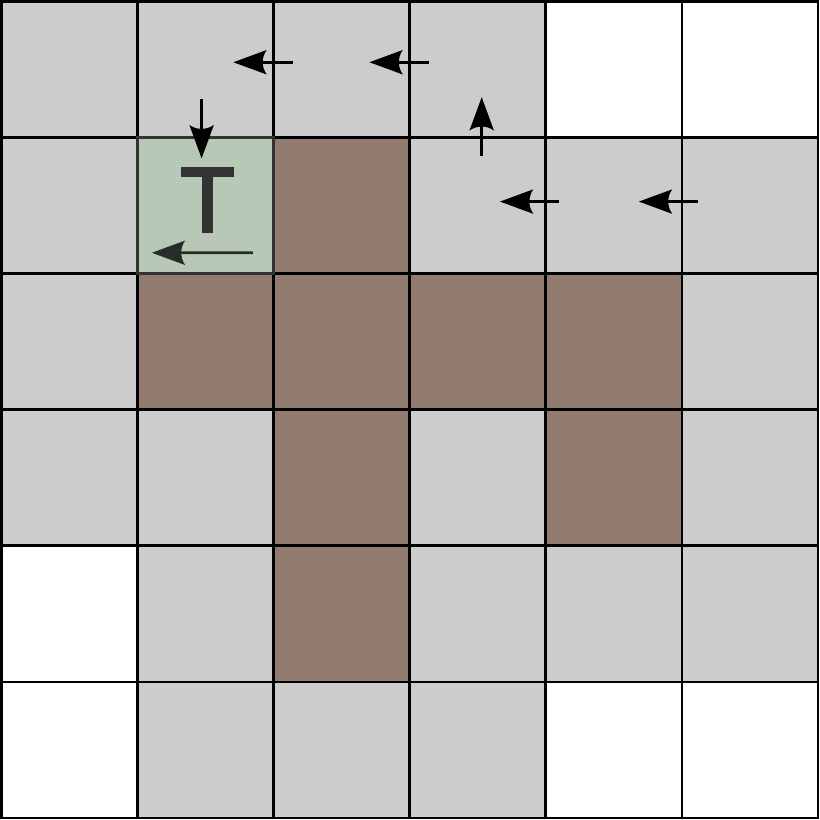}
\caption{A token traversing the surface of a configuration. The surface of the configuration is denoted by light grey tiles. The cell labeled $T$ is in $\mathtt{token\_state\_left}$ as indicated by the arrow depicted on the cell.}
\label{fig:walkingToken}
\end{center}
\end{figure}

The goal of this construction is to give a synchronous nondeterministic CA $\mathcal{B} = (\mathbb{Z}^2, S, N, \delta)$ and initial configuration that simulates an arbitrary aTAM system $\mathcal{T} = (T,\sigma, \tau)$. The neighborhood of $\mathcal{B}$ is the Moore neighborhood and
the states and local rules for $\mathcal{B}$ are obtained as follows. First, we add a state to $S$ for each tile type of $\mathcal{T}$ we call these states $\mathtt{tile\_states}$.  We also add states $\mathtt{token\_state\_up}$,
$\mathtt{token\_state\_left}$, $\mathtt{token\_state\_down}$ and $\mathtt{token\_state\_right}$ and we use $\mathtt{token\_states}$ to refer to any of these $4$ states.
We refer to any cell in a $\mathtt{token\_state}$ as the {\em token}. This token moves one cell counterclockwise at each time step and only one cell is in a $\mathtt{token\_state}$ at any given time.
At each time step, the cell in a $\mathtt{token\_state}$ moves one cell either up, left, down or right in an effort to traverse the {\em surfaces} of an existing
configuration, where a surface of a configuration is the connected set of quiescent cells
that neighbor (using the Moore neighborhood) at least one non-quiescent state. (See Figure~\ref{fig:walkingToken}.) Note that a configuration may have many disjoint surfaces.

\begin{figure}[htp]
\begin{center}
\includegraphics[width=3in]{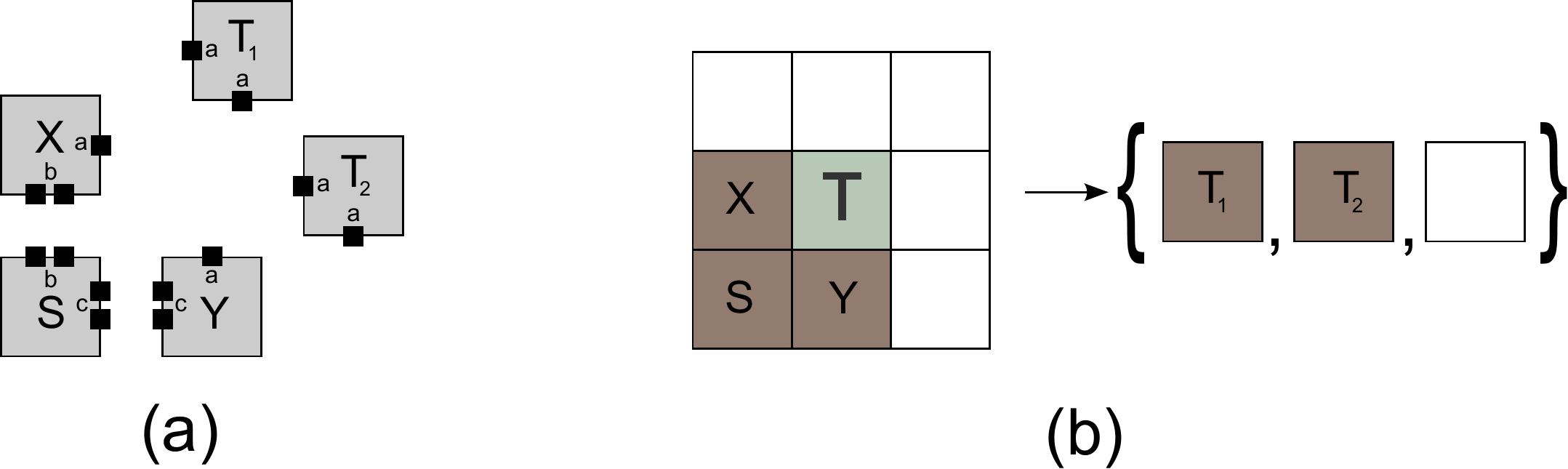}
\caption{{\bf (a)} A tile set consisting of $5$ tile types. Glues $b$ and $c$ have strength $2$ and $a$ glues have strength $1$. {\bf (b)} A local rule corresponding to the tile set in {\bf(a)}. Cells in the Moore neighborhood that are in $\mathtt{tile\_states}$ are labeled with the label of the corresponding tile type in {\bf (a)}. The cell labeled $T$ is in a $\mathtt{token\_state}$. Blank cells are quiescent.}
\label{fig:ruleFromTileSet}
\end{center}
\end{figure}

 At each time step, a cell in a $\mathtt{token\_state}$ can nondeterministically transition to a $\mathtt{tile\_state}$ if and only if the tile corresponding to this state could bind in the simulated aTAM system. This ensures that at any given time step, at most one cell transitions from a quiescent state to a $\mathtt{tile\_state}$. Figure~\ref{fig:ruleFromTileSet} shows an example of a local rule obtained from a tile set.

\begin{figure}[htp]
\begin{center}
\includegraphics[width=1.2in]{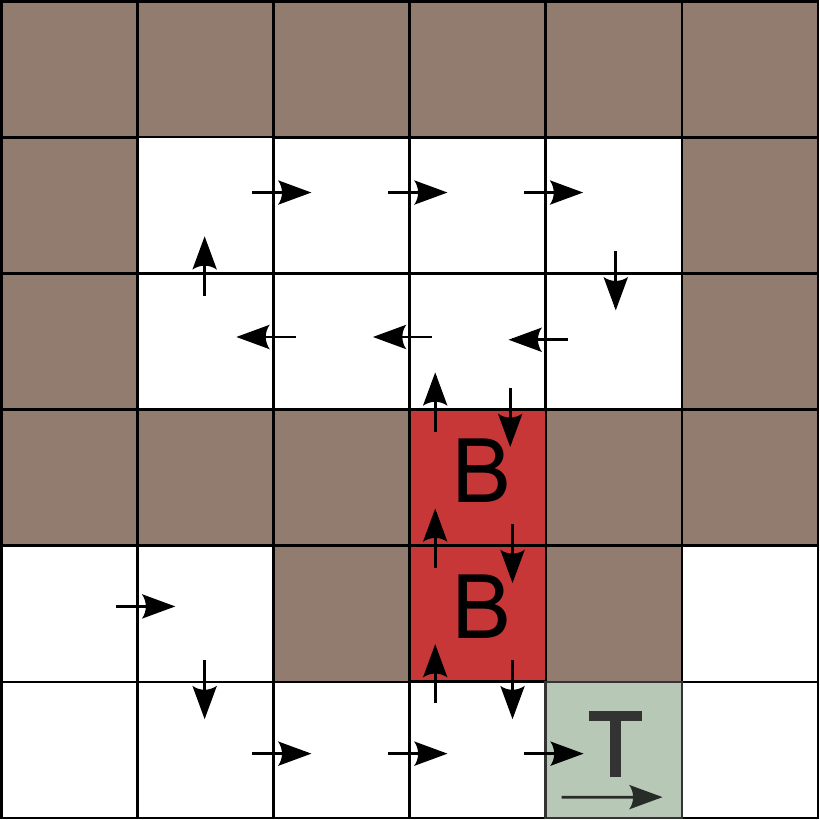}
\caption{Traversing two disjoint surfaces using a single token and $\mathtt{bridge\_tile\_states}$}
\label{fig:tokenTracing}
\end{center}
\end{figure}

The idea is that the token can put cells in a $\mathtt{tile\_state}$ on the surface of a configuration. However a configuration may have many disjoint surfaces.
Non-quiescent states of a configuration can break the $\mathbb{Z}^2$ lattice into disjoint regions of cells in quiescent states. For example, this can occur when the CA is simulating a tile set that assembles a frame, i.e. tiles around some square of empty tiles.
This leads to disjoint surfaces that the token must traverse. Therefore, care must be taken in order to allow the token to traverse surfaces separated by non-quiescent states. This is accomplished by adding a $\mathtt{bridge\_tile\_state}$ to $S$ for each tile type of the simulated aTAM system. The token is allowed to ``pass over'' these $\mathtt{bridge\_tile\_states}$. Passing over a cell in $\mathtt{bridge\_tile\_state}$ is done by adding a $\mathtt{bridge\_tile\_token\_state}$ to $S$ for each tile type in $T$ and each direction $\mathtt{up}$, $\mathtt{left}$, $\mathtt{down}$ and $\mathtt{right}$.
Figure~\ref{fig:tokenTracing} shows the token and its path as it traverses two surfaces of the configuration by crossing $\mathtt{bridge\_tile\_states}$ . When transitioning to a $\mathtt{tile\_state}$ or $\mathtt{bridge\_tile\_state}$, we can determine which state to transition to by using Moore neighborhoods. For more details on how $\mathtt{token\_states}$ or $\mathtt{bridge\_tile\_states}$ work, see Section~\ref{sec:tokenBridgeDets}.

\later{
\section{Construction details CA simulation of aTAM system}\label{sec:casimtasdets}
\subsection{Local rules involving the token and bridge tile states}\label{sec:tokenBridgeDets}
Here we describe the local rules that allow the token to traverse the surface of a configuration and how cells may transition to a $\mathtt{bridge\_tile\_state}$.
The direction of each $\mathtt{token\_state}$ determines its future direction of ``movement'' so that a Moore neighborhood can be used to determine the direction of travel. Specifically, the direction of the state
refers to the relative position of a quiescent cell that will change to the a $\mathtt{token\_state}$. Figure~\ref{fig:tokenRuleExample} shows a local rule for a neighborhood with a token.
\begin{figure}[htp]
\begin{center}
\includegraphics[width=3in]{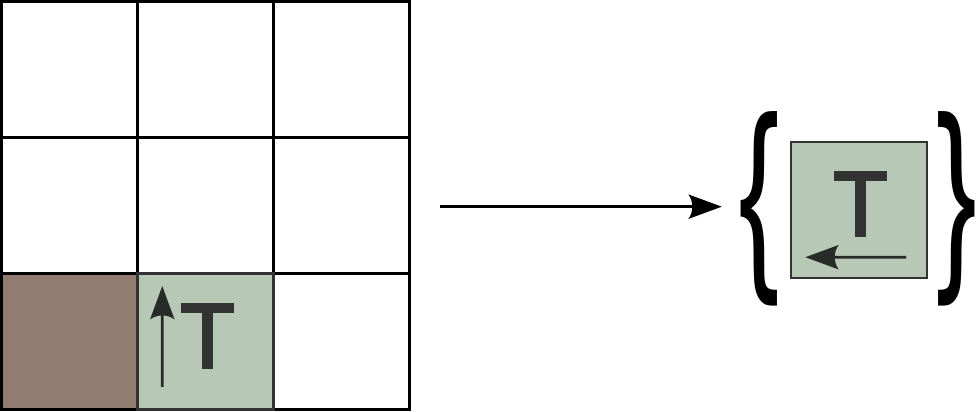}
\caption{In the Moore neighborhood, we can determine the direction of the next $\mathtt{token\_state}$.}
\label{fig:tokenRuleExample}
\end{center}
\end{figure}

To understand the need for $\mathtt{bridge\_tile\_states}$, notice that with our CA, all points in $\mathbb{Z}^2$ that map to non-quiescent states are neighbors in the lattice. 
This follows from the fact that only the token can transition to a $\mathtt{tile\_state}$. 
 In this case, we say that the configurations of the CA are {\em connected}. Figure~\ref{fig:mooreNeighborhood} shows a Moore neighborhood of a cell in a $\mathtt{token\_state}$.
\begin{figure}[htp]
\begin{center}
\includegraphics[width=1.2in]{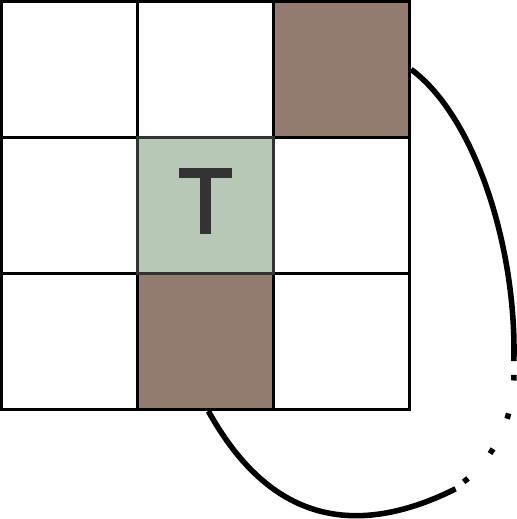}
\caption{Transitioning the center cell to a $\mathtt{tile\_state}$ would divide the neighborhood's quiescent states. Therefore the center cell transitions to a $\mathtt{bridge\_tile\_state}$}
\label{fig:mooreNeighborhood}
\end{center}
\end{figure}
Under the condition
that the configuration is connected we examine the $8$ cells of the Moore neighborhood around the center cell. If transitioning to a $\mathtt{tile\_state}$ results in dividing the quiescent points of the lattice restricted to the neighborhood into two disjoint subsets,
the center cell transitions to a $\mathtt{bridge\_tile\_state}$, otherwise it transitions to a $\mathtt{tile\_state}$. The algorithm to do this treats a cell already in $\mathtt{bridge\_tile\_state}$ like one in the quiescent state. Figure~\ref{fig:bridgeTileState} depicts a
time step where a cell in a $\mathtt{token\_state}$ transitions to a cell in a $\mathtt{bridge\_tile\_state}$.  Note that transitioning to a $\mathtt{tile\_state}$ would ``trap'' the token, but transitioning to a $\mathtt{bridge\_tile\_state}$ allows the token to traverse multiple surfaces.
\begin{figure}[htp]
\begin{center}
\includegraphics[width=3in]{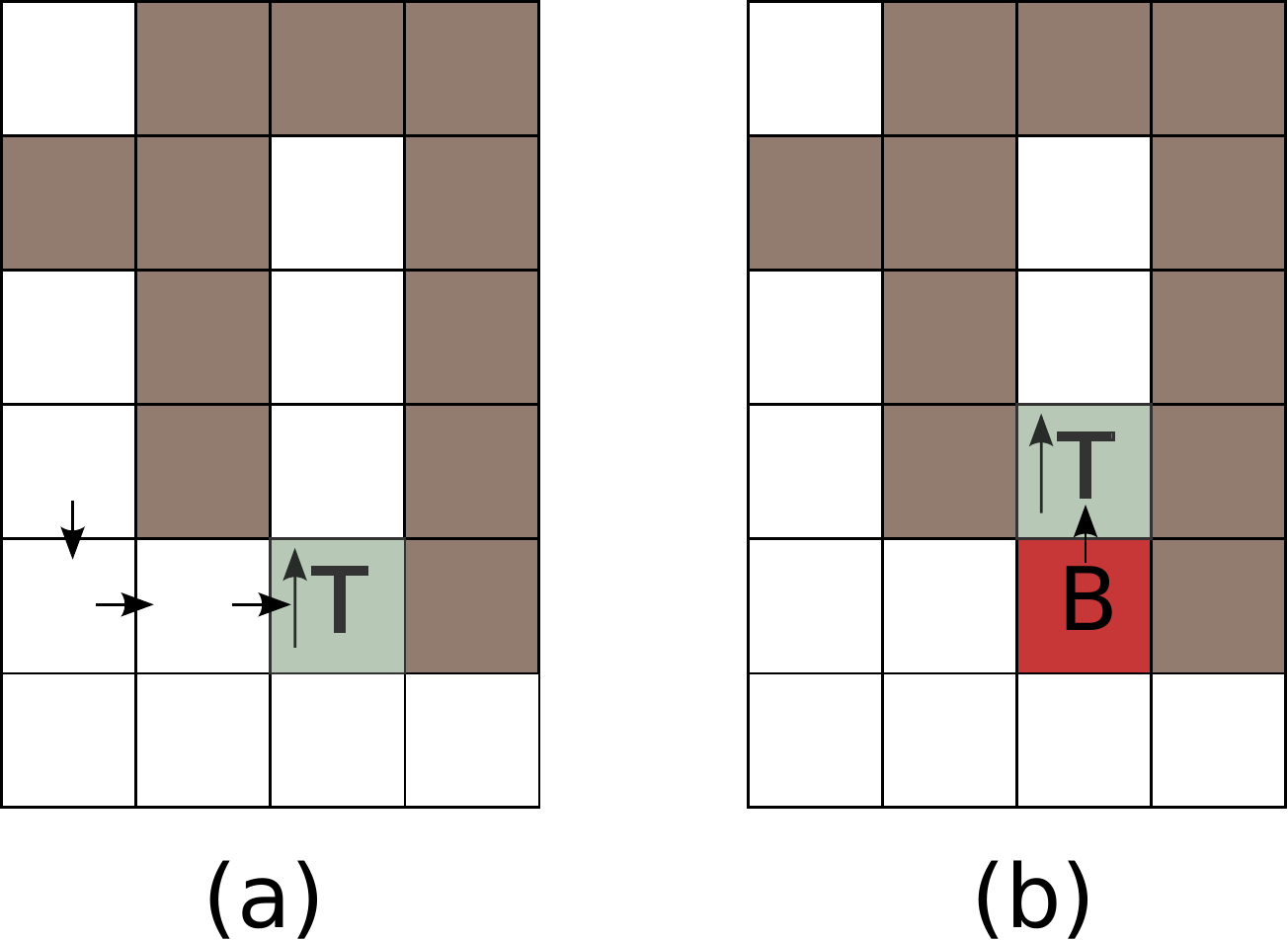}
\caption{{\bf(a)} A $\mathtt{token\_state}$ prior to transitioning to a $\mathtt{bridge\_tile\_state}$ {\bf (b)} The surface is split into two surface. The token must pass over a cell in $\mathtt{bridge\_tile\_state}$ to completely traverse both surfaces.}
\label{fig:bridgeTileState}
\end{center}
\end{figure}
}
If and when a path of cells in a $\mathtt{bridge\_tile\_state}$ no longer leads to quiescent states and the final quiescent state transitions to a $\mathtt{tile\_state}$, the token traverses the cells in $\mathtt{bridge\_tile\_states}$ as it continues its counterclockwise traversal of a configuration. Since the path of cells in $\mathtt{bridge\_tile\_states}$ no longer leads to any quiescent states, as the token traverses cells in $\mathtt{bridge\_tile\_states}$, these cells transition to $\mathtt{tile\_states}$ that
correspond to their $\mathtt{bridge\_tile\_state}$ counterparts. As a result, the token no longer unnecessarily traverses a path of cells in $\mathtt{bridge\_tile\_states}$ that would only lead to other cells in $\mathtt{bridge\_tile\_states}$.

\begin{figure}[htp]
\begin{center}
\includegraphics[width=3in]{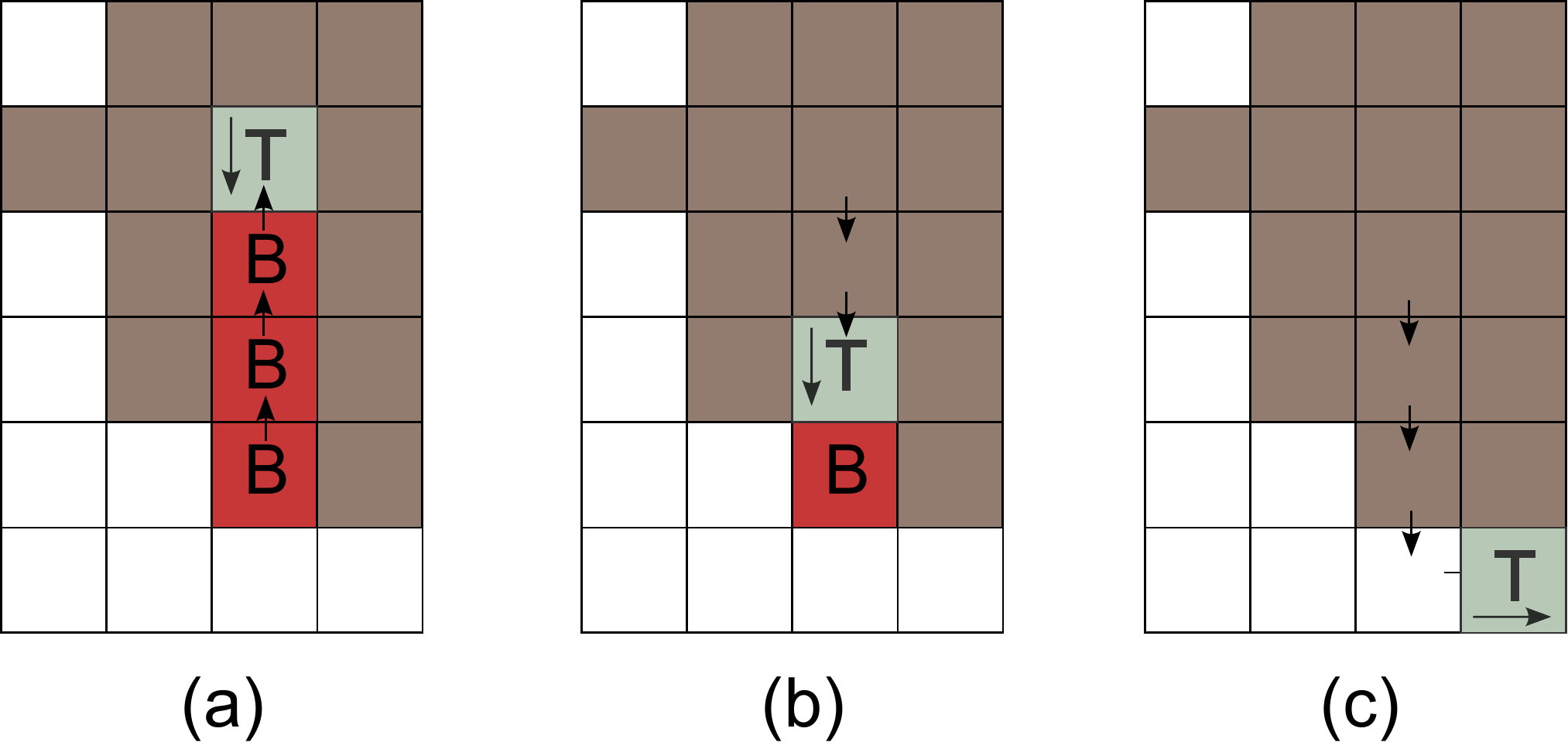}
\caption{{\bf (a)}The token prior to transitioning to a $\mathtt{tile\_state}$. {\bf(b)} Two time steps later we see a cell in $\mathtt{bridge\_tile\_state}$ has transitioned to just a $\mathtt{tile\_state}$. {\bf(c)} The entire path of cells in $\mathtt{bridge\_tile\_states}$ have transitioned to corresponding $\mathtt{tile\_states}$.}
\label{fig:collapsingBridges}
\end{center}
\end{figure}

An example of a CA simulating an aTAM system can be found at {\url{http://self-assembly.net/CASimTAS}}. There are also instructions for creating a CA based on an aTAM system.

\later{
\subsection{Proof of Correctness}\label{sec:pocCAsimTAS}
Let $\mathcal{T} = (T,\sigma, \tau)$ be an aTAM system. First we show that given an aTAM system, the construction in Section~\ref{sec:caconstruction} can be used to give a CA that simulates
$\mathcal{T}$. Let $\mathcal{A}$ be the cellular automaton obtained by the construction. In other words, let $\mathcal{A}$ is the CA with states $\mathtt{tile\_states}$, $\mathtt{bridge\_tile\_states}$ and $\mathtt{bridge\_tile\_token\_states}$ corresponding to tile types of $\mathcal{T}$ as well as $4$ $\mathtt{token\_state}$.
We can take the rescaling $\mathcal{A^{\prime}}$ to be
the trivial rescaling of $\mathcal{A}$. In other words, we take $\mathcal{A^{\prime}}$ to just be $\mathcal{A}$.
Then we take the representation function $R$ to be the partial function that maps a cell with state $\mathtt{tile\_states}$, $\mathtt{bridge\_tile\_states}$ or $\mathtt{bridge\_tile\_token\_states}$ to a tile with
tile type that corresponds to the state representing this particular tile type.
The initial configuration $c_0$ of $\mathcal{A}$ can be obtained from $\sigma$ by first
mapping each point in $\dom{\sigma}$ to a the corresponding $\mathtt{tile\_states}$. Then, since in general non-quiescent cells must be connected, but could divide quiescent cells into disjoint regions of the lattice,
we connect any disjoint regions of connected quiescent cells by paths of non-quiescent cells. Diagonal quiescent cells are not considered connected. Then we change the states of the cells along this path to corresponding $\mathtt{bridge\_tile\_states}$. Figure~\ref{fig:initConfig} gives an example of changing changing states along paths connecting disjoint regions of quiescent cells to $\mathtt{bridge\_tile\_states}$. Finally, we put a
cell just above a cell in a $\mathtt{tile\_state}$ in $\mathtt{token\_state\_left}$.

\begin{figure}[htp]
\begin{center}
\includegraphics[width=4.5in]{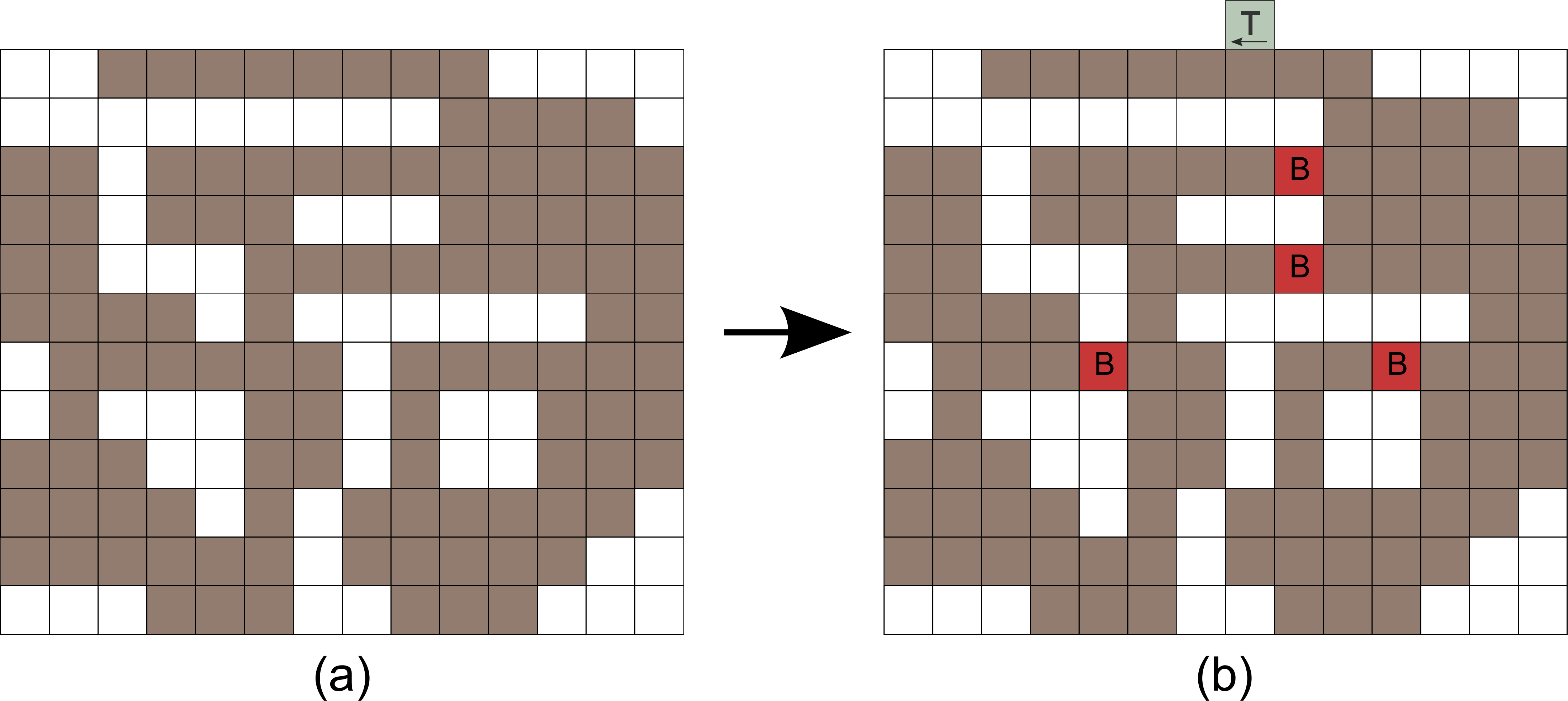}
\caption{An initial configuration obtained from an initial assembly before {\bf(a)} and after {\bf(b)} changing states along paths connecting disjoint regions of quiescent cells to $\mathtt{bridge\_tile\_states}$. Grey cells denote cells in any $\mathtt{tile\_state}$ while cells labeled $B$ denote cells in $\mathtt{bridge\_tile\_states}$. The cell labeled $T$ is in $\mathtt{token\_state\_left}$.}
\label{fig:initConfig}
\end{center}
\end{figure}

Now, to see that $\left(\mathcal{A}, c_0\right)$ simulates $\mathcal{T}$, notice that the token enforces that at most a single cell of a configuration of $\mathcal{A}$ transitions to
$\mathtt{tile\_state}$, $\mathtt{bridge\_tile\_state}$ or $\mathtt{bridge\_tile\_token\_state}$. Therefore applying the global rule to a configuration of $\mathcal{A}$ results in a configuration where either the token has moved, or the token has moved and a quiescent state transitions to a state representing a tile. In the former case, the configuration before and after application of the global rule represent the same assembly. In the latter case, letting $\alpha$ be the assembly represented by the configuration $c$ prior to applying the global rule we see that any configuration in $G(c)$ represents an assembly obtained by adding a single tile to $\alpha$ of the necessary type.
In other words, single state changes from the quiescent state to a state representing a tile type correspond to additions of single tiles in the aTAM system. Hence, $\mathcal{T}$ {\em follows} $\left(\mathcal{A}, c_0\right)$. Likewise, any possible single tile binding in $\mathcal{T}$ corresponds to some possible transition of a
quiescent state to a state that represents the tile type of the binding tile in the corresponding location (with perhaps several transitions which only move the token into the correctly corresponding location). Therefore, $\left(\mathcal{A}, c_0\right)$ {\em models} $\mathcal{T}$. This shows Lemma~\ref{lem:casim}.

To show Theorem~\ref{thm:caiu}, let $\mathcal{T}$ be an arbitrary aTAM system and let $\mathcal{U} = (U,\sigma_{\mathcal{T}},2)$ be the aTAM system that uses the tile set of \cite{IUSA}, which is intrinsically universal for the aTAM, to simulate $\mathcal{T}$ under $R_{t}$.  Now let $\mathcal{A}$ be the CA which simulates $\mathcal{U}$ under $R_{a}$ as in the proof of Lemma~\ref{lem:casim}.
Then note that $\sigma_{\mathcal{T}}$ gives rise to an initial configuration $c_0$ of $\mathcal{A}$. With this initial configuration, $R_{a}^*$ followed by $R_{t}^*$ maps configurations of $\mathcal{A}$ with initial configuration $c_0$ to assemblies of $\mathcal{T}$. This composition of maps gives a representation function that shows that $\left(\mathcal{A}, c_0\right)$ simulates $\mathcal{T}$.

}

\section{An aTAM Tile Set Which Can Simulate Any Nondeterministic CA}\label{sec:TAsimCA}

\begin{theorem}\label{thm:TAMsimIU}
There exists an aTAM tile set $U$ which is able to simulate the entire class of nondeterministic CA systems with finite initial configurations.  %
\end{theorem}

Theorem~\ref{thm:TAMsimIU} states that there is a single tile set $U$ in the aTAM which can be used to form a TAS $\mathcal{U} = (U,\sigma_{\mathcal{C}},2)$ which is dependent upon a given CA $\mathcal{C}$, for any arbitrary nondeterministic CA $\mathcal{C}$ and a finite initial configuration for $\mathcal{C}$, where the seed to the aTAM system encodes information about $\mathcal{C}$ and its initial configuration, so that $\mathcal{U}$ simulates $\mathcal{C}$.  In order to prove Theorem~\ref{thm:TAMsimIU}, we will progress in two steps, first proving the following Lemma.

\begin{lemma}\label{lem:TAMsimGOL}
Let CA $\mathcal{A}$ be Conway's Game of Life.  There exists an aTAM tile set $U$ and a scalable representation function $R$ such that, given $c_0$ as an arbitrary but finite initial configuration of $\mathcal{A}$, there exists an aTAM TAS $\mathcal{T} = (U, \sigma_{c_0}, 2)$ such that $\mathcal{T}$ simulates $\left(\mathcal{A}, c_0\right)$.
\end{lemma}

Lemma~\ref{lem:TAMsimGOL} states that there exists a single tile set $U$ in the aTAM which can be used to simulate the Game of Life CA given any finite initial configuration.  We now present a construction to prove this.

\subsection{Overview of construction to prove Lemma~\ref{lem:TAMsimGOL}}

The system $\mathcal{T} = (U,\sigma_{c_0},2)$ will be designed so that the seed is a single line of tiles which encodes the initial configuration, $c_0$, of $\mathcal{A}$.  Assume that all non-quiescent cells within $c_0$ can fit into an $n \times n$ square. (Throughout this discussion, we will refer to a \emph{cell} as exactly one of the cells of $\mathcal{A}$ and the \emph{grid} as the full set of cells being simulated at a given time.  A \emph{step} refers to a single time step of $\mathcal{A}$ and a \emph{stage} refers to the assembly representing the entire grid at a particular step.)  The encoding of the initial configuration consists of a listing of the states of each of the $n^2$ cells within that box.  Since it is possible that, at each time step $0<t<\infty$, a cell which was previously quiescent and which was just outside the boundary of the currently simulated grid switches its state to a non-quiescent value, to accurately simulate the full behavior of $\mathcal{A}$ we must simulate an increasingly larger grid at each time step.  In order to assure that no (non-quiescent) behavior of $\mathcal{A}$ could occur beyond the bounds of our simulation, at each stage we increase the dimensions of the grid by $2$, adding a row of cells to each of the top and bottom, and a column to each of the left and right.  We say that we perform a recursive, ``in-place'' simulation of $\mathcal{A}$, namely one in which every subassembly which maps to a single cell at some time step $t$ contains within it, at smaller scale factors, the entire configuration of $\mathcal{C}$ at \emph{every} time step $t'<t$ (recursive), and also that the subassembly mapping to any single cell at any time step $t$ is contained within an infinite hierarchy of subassemblies which each map to a unique cell at some time step $t'$ where $t < t' < \infty$, i.e. each simulated cell and grid is fully encapsulated within the simulation of a single cell at the next greater time step (in-place).

\begin{figure}[htp]
\begin{center}
\includegraphics[width=3.5in]{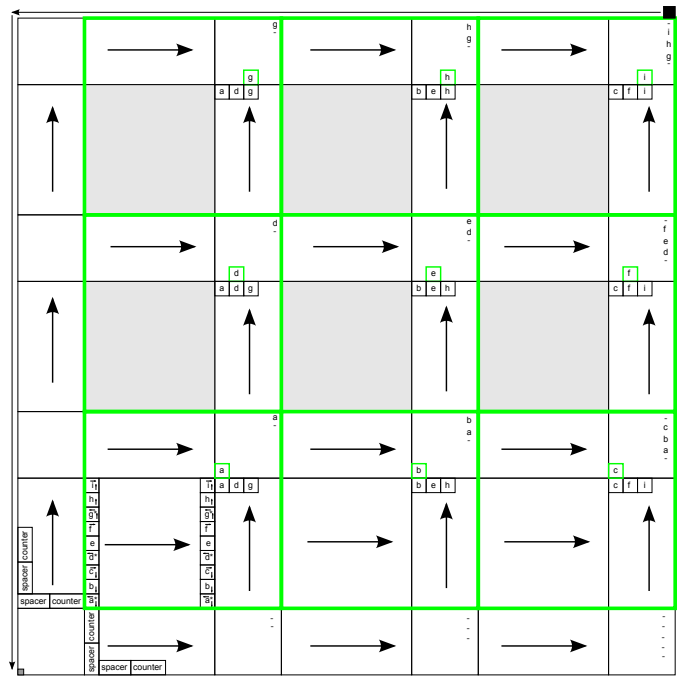}
\caption{Completed formation of the representation of time step $0$.}
\label{fig:TASsimCA-stage1done}
\end{center}
\end{figure}

\begin{figure}[htp]
\begin{center}
\includegraphics[width=5.0in]{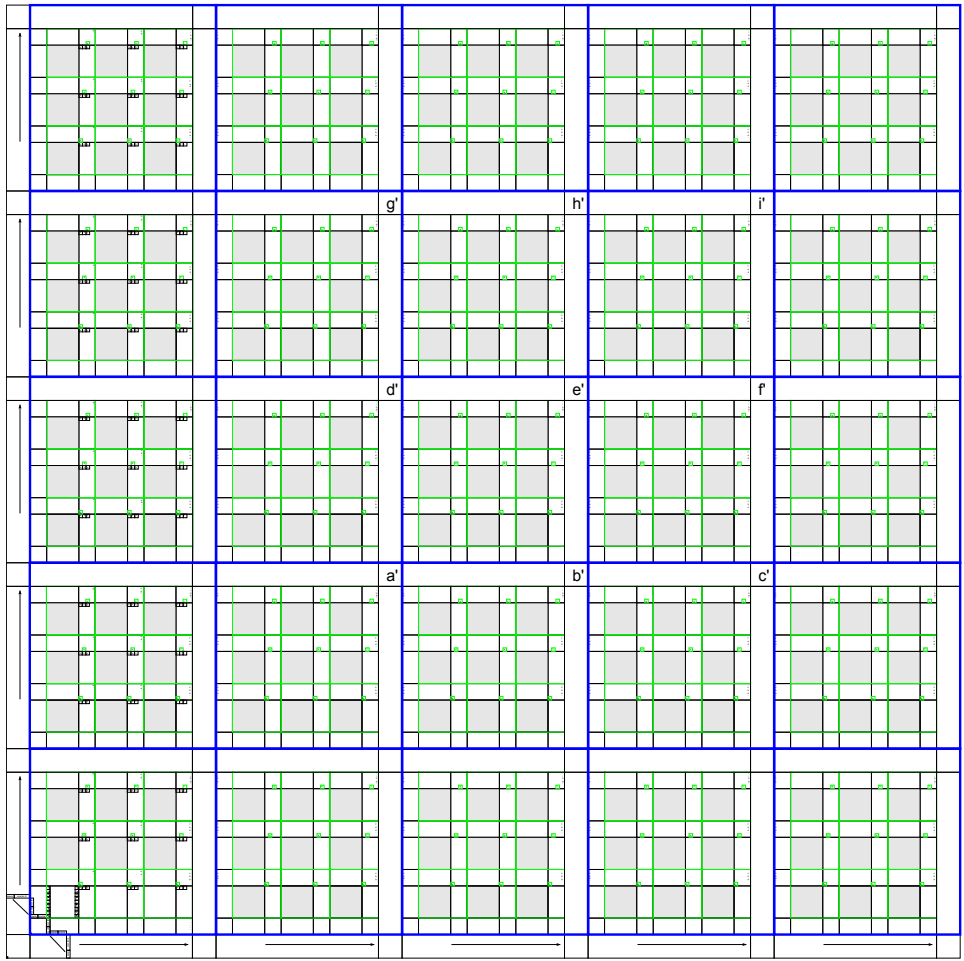}
\caption{Completed growth of the second stage.  Blue squares depict the representations of the cells of that stage.}
\label{fig:TASsimCA-stage2done}
\end{center}
\end{figure}

See Figures~\ref{fig:TASsimCA-stage1done} and \ref{fig:TASsimCA-stage2done} for high-level depictions of the simulation of time steps $0$ (the initial configuration of $\mathcal{A}$) and $1$ (the first transition of $\mathcal{A}$).  Details of the construction can be found in Section~\ref{sec:TASsimCA-append}.

\vspace{-10pt}

\later{
\section{Construction details for aTAM simulation of CA}\label{sec:TASsimCA-append}

To show how our construction works, we will break it down into a series of modules, describing how each works independently, and then describe how they are all combined for the full construction.  (Note that these modules consist mainly of tiles which assemble to perform well-known primitives, such as counting and rotating values, used in various aTAM constructions.  Therefore, many of the details of these modules are omitted except where relevant modifications are made to the commonly used versions.  The reader is encouraged to see \cite{IUSA} for additional details about many such primitives.)

Note that $\mathcal{A} = (\mathbb{Z}^2, S, N, \delta)$ uses the Moore neighborhood, i.e. $\{(x,y)|x,y\in\{-1,0,1\}\}$, and that $S = \{0,1\}$.  Let $n \in \mathbb{N}$ be the dimensions of a square bounding box which completely encircles all cells in $c_0$ which are not quiescent, along with at least one ring of quiescent cells around the perimeter,
then let $c'_0$ be the $n \times n$ square of cells contained within that box.  From here onward, we will refer to $c'_0$ as the initial configuration for the CA system being simulated.

\subsection{Seed configuration}

We construct $\sigma_{c_0}$, the seed assembly for $\mathcal{T}$, as a single column of tiles.  Let $n$ be the dimensions of $c'_0$, and note that it requires $n^2$ tiles to encode $c'_0$.  The eastern glues of $\sigma_{c_0}$, starting from the bottom and moving up, encode $c'_0$ by encoding the states of the bottom row of cells in $c'_0$ from left to right, followed by the row in $c'_0$ directly above that, etc., fully encoding the $n \times n$ block of cells in $c'_0$ in a line.  For each position representing a cell which is the leftmost cell in its row of $c'_0$, a special `$*$' marker is added, and for each border of the $n\times n$ block which the cell is on, a corresponding arrow marker is added. An example can be seen in Figure~\ref{fig:TASsimCA-seed}.

\begin{figure}[htp]
\begin{center}
\includegraphics[width=1.5in]{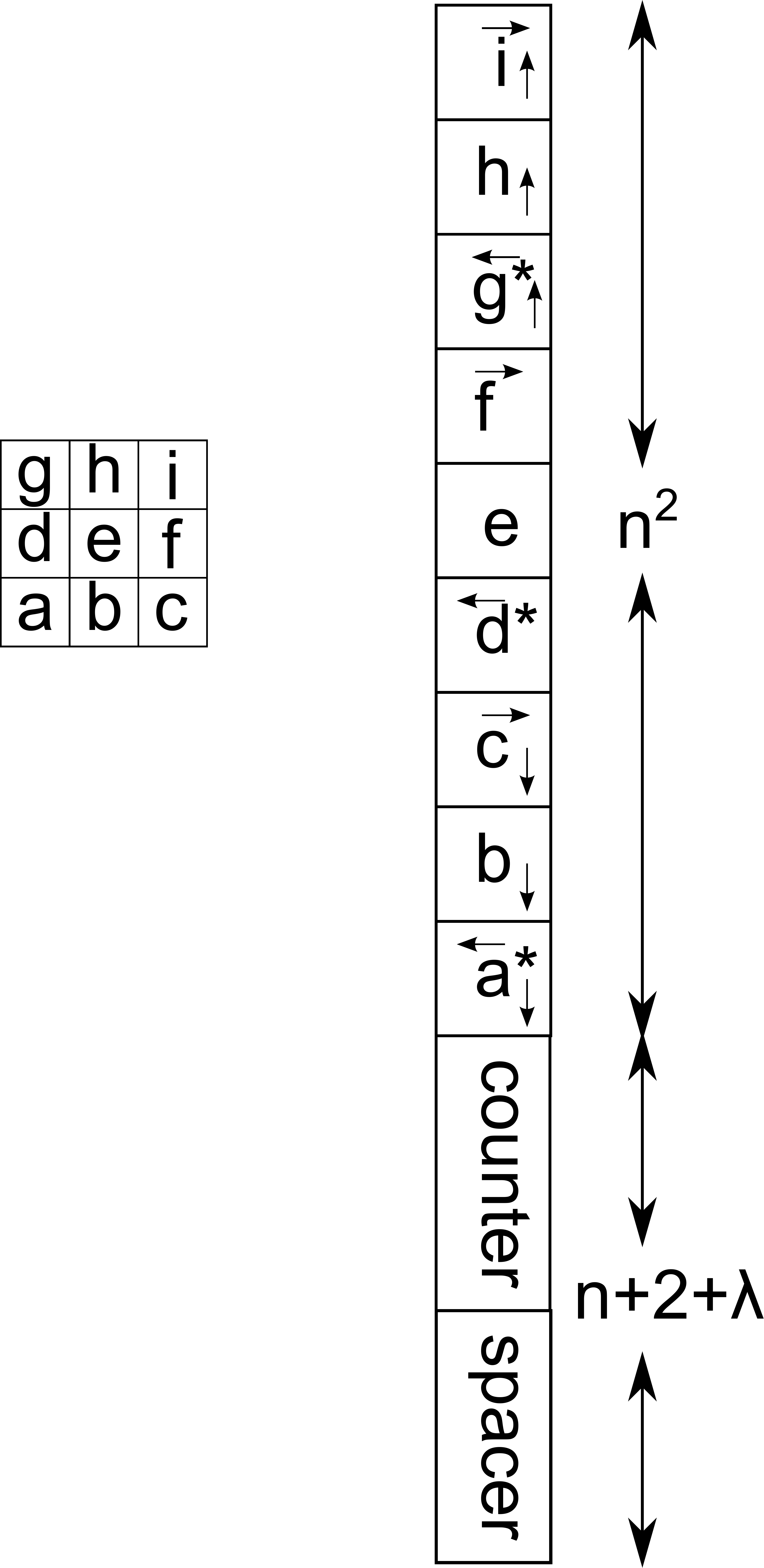}
\caption{(Left) An example $3\times3$, or $n=3$, initial configuration $c'_0$ with each cell given a uniquely identifying letter for identification purposes, (Right) The row of tiles encoding $c'_0$, showing the ordering of the representation of cells in relation to $c'_0$.}
\label{fig:TASsimCA-seed}
\end{center}
\end{figure}

Below the encoding of $c'_0$ are the portions of the seed assembly labeled ``counter'' and ``spacer'' in Figure~\ref{fig:TASsimCA-seed}.  The ``counter'' portion contains the number $n-1$ encoded in binary which will be used as the maximum value for assembling counters, and the ``spacer'' portion simply contains some spacer tiles which are used to ensure that the full height of the ``counter'' + ``spacer'' portions is equal to $n + 2 + \lambda$, where $\lambda$ is equal to 2 for this version of the construction.

\subsection{Initial seed growth}\label{sec:seedGrowth}

The growth from the seed column $\sigma_{c_0}$ occurs in three ways (which are all represented in Figure~\ref{fig:TASsimCA-seed-to-block}).  The portion which grows to the east along the bottom (beginning from the ``counter'' section) contains a set of $3$ nested binary counters, $ctr_0$, $ctr_1$, and $ctr_2$, which each start at $0$ and count to a maximum value of $n-1$ as follows:  $ctr_0$ increments at every column, $ctr_1$ increments every time $ctr_0$ reaches the value $n-1$ (at which point $ctr_0$ resets to $0$ and continues counting), and $ctr_2$ increments every time $ctr_1$ reaches the value $n-1$.  In this way, once $ctr_1$ has counted to $n-1$, the full length traveled is $n^2$.  This type of growth is possible since the counter grows in a standard zig-zag manner, and it passes forward the encoded value of $n-1$ which it uses to compare against current counter values.  Additionally, while this segment is growing to the east, it also rotates downward the information encoded in its initial row (i.e. the maximum counter value and spacing).  The north surface of the counter provides a base across which the representation of $c'_0$ grows, to the east.  The west side of the seed column initiates growth which rotates the counter and spacer values clockwise, while forming a square with a specially marked corner (shown as grey in Figure~\ref{fig:TASsimCA-seed-to-block}) for reasons to be discussed later.  The north facing counter then grows in the same way as the east facing counter.  The southern counter grow at a width of $n+2+\lambda$, and the western counter's width is $n+1+\lambda$, which are both much wider than necessary for the counters themselves, for reasons which will be discussed later.  Once they have reached the distance $n^2$, the counters ``pause'' counting while a square of dimension $n+2+\lambda$ form.

\begin{figure}[htp]
\begin{center}
    {\subfloat[{\scriptsize Counters grow along the south and west, while the values of $c'_0$ move east.}]
    {\label{fig:TASsimCA-seed-to-block}\includegraphics[height=2.3in]{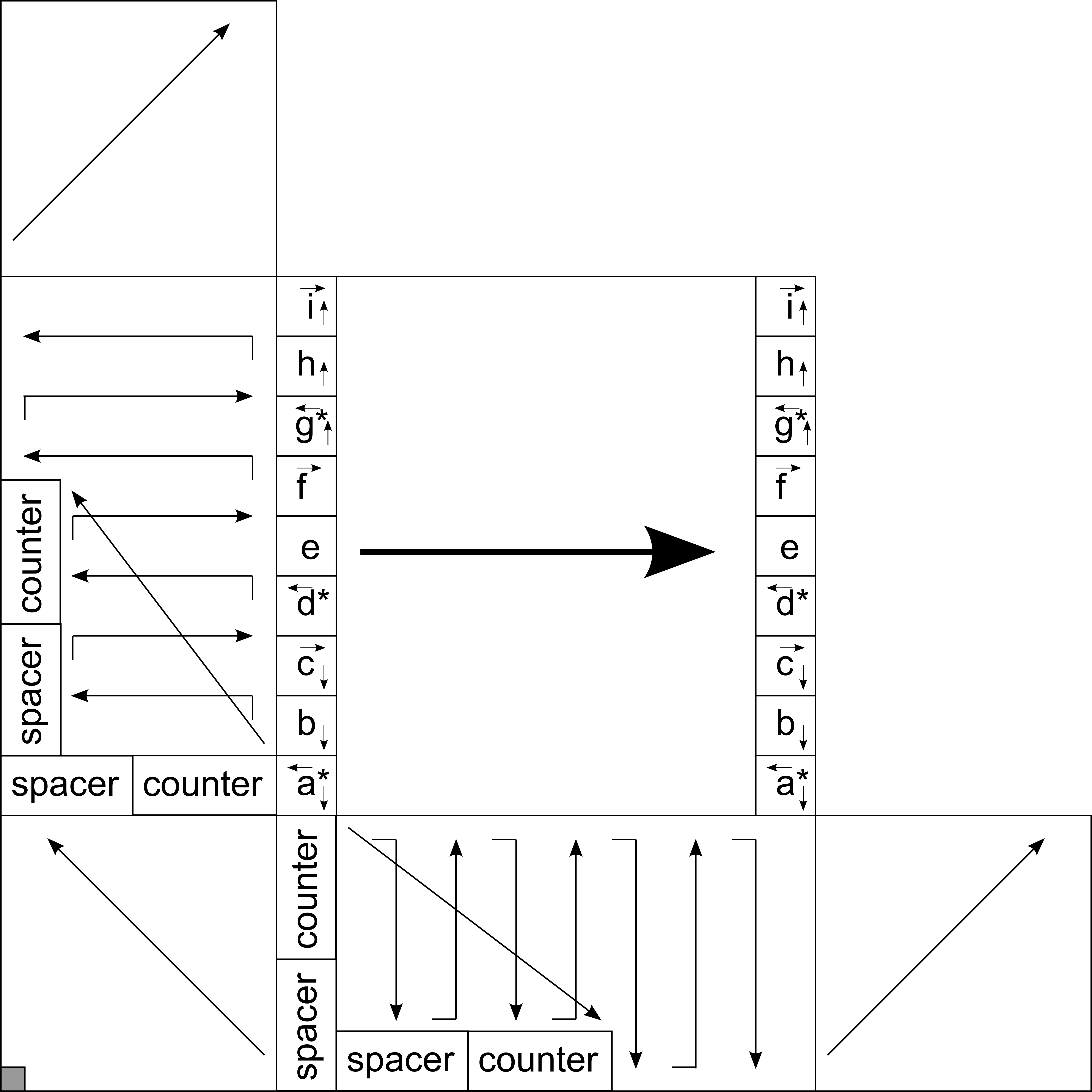}}}
    \quad\quad
    {\subfloat[{\scriptsize Subsequent growth which ``selects'' the values from the first column of $c'_0$.}]
    {\label{fig:TASsimCA-seed-to-block2}\includegraphics[height=2.3in]{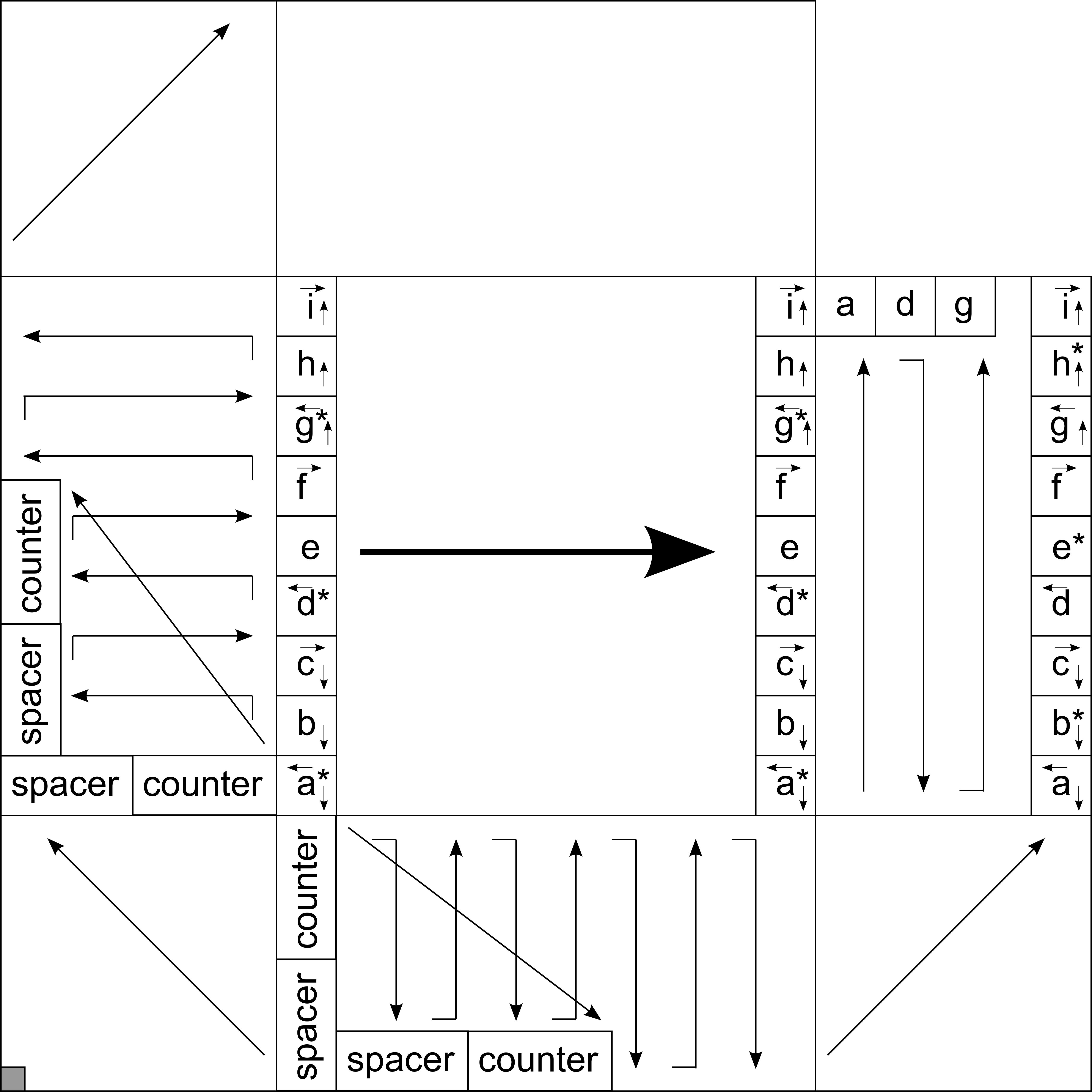}}}
    \caption{The initial growth of the seed assembly shown in Figure~\ref{fig:TASsimCA-seed}.}
\end{center}
\end{figure}

\subsection{Completion of initial stage}

The box on the bottom right of Figure~\ref{fig:TASsimCA-seed-to-block} initiates the growth of a series of zig-zag, upward and downward growing columns which 1) propagate the encoding of $c'_0$ to the right while shifting the ``$*$'' marks up by one position (thus marking the locations of the cells in the second column), and 2) copy all values with ``$*$'' marks (before the shift) to the top row.  Note that these are guaranteed to fit with no more than one value per column since the width of the rectangle is that of the outer counters which is $n+2+\lambda$ and there are only a total of $n$ values for any row.  The square on top of the western counter initiates the growth of a rectangle which reaches the location where the remaining values for the first column of $c'_0$ (i.e. $d$ and $g$ in the example) are located.  This allows a square to form which selects the first value (in the example, $a$) as the value representing this simulated cell.

At this point, the southern and western counters are both able to continue growing, with their $ctr_0$ and $ctr_1$ values reset to $0$ while their $ctr_2$ values are both incremented to $1$.  Further growth similar to the pattern of growth up to this point continues until the $ctr_2$ counters have each reached their maximum value, resulting in the full formation of the assembly representing the initial configuration of $\mathcal{A}$, as shown in Figure~\ref{fig:TASsimCA-stage1done}.  It can be seen how, at this point, the values for each of the grid locations in $c'_0$ have been selected in regions corresponding to their locations in $c'_0$ (highlighted in green) by performing simple shifting of markers across the values propagated through the \emph{fibers}.

We refer to the rectangular portions and the smaller squares as \emph{fiber}.  The fibers on the western and southern sides of the assembly of any given stage are called \emph{boundary fibers}, while the rest are called \emph{stage fibers}.  The bottom row of upward growing stage fibers perform the column selection and shifting of the ``$*$'' marks as previously discussed.  The upper rows simply select the leftmost marked values which are passed to them and propagate all remaining values upward while shifting the mark by one position.  Each horizontal fiber collects all values for its row, picking them up one at a time from left to right, getting one from each square where it crosses with a vertical fiber. See Figure~\ref{fig:TASsimCA-horiz-fiber} for an example.  It does this by designating each of its rows to carry the value of one cell in that row, with the top and bottom hardcoded to each represent a cell in the quiescent state (and note that when these are added they are initialized to contain the necessary markers denoting whether they are at the left or right side of the grid - see Figure~\ref{fig:TASsimCA-seed} for an example of the arrow markers).  Since it will gather the $n$ cell values for the current stage, the addition of a quiescent value to the bottom and top simulates the addition of an extra cell on the left and right side of this row in the grid for the next stage (since this listing of cell values will be used to compute the values for the next stage), and the height of this fiber is $n+2+\lambda$, ensuring that all values will fit with one per row.  The horizontal fiber grows in a zig-zag manner, using the square below it to guide it until it reaches an intersection with a vertical fiber.  Through the square of the intersection, the columns grow just in the upward direction, propagating all of the information about the cell values for that column upward while using cooperation from the west to propagate the ``$*$'' marker one position to the right, and to also collect that value all the way to the top right of the square.  After leaving the intersection, it resumes its zig-zag growth, which allows it to shift all collected cell values for that row down by one (other than the top and bottom values, which stay in fixed positions).

The horizontal fiber provides for a portion of the addition of a ring of quiescent cells for the next stage.  To handle the addition of the new bottom row, the boundary fiber creates a similar list of values, with all set to quiescent and containing the necessary markings for the edges of the grid that they are on.  To handle the addition of another row on the top, the next stage will insert quiescent values during the distribution of cell values.  The squares at the intersections of stage fibers all grow using cooperation between the two fibers, in order to preserve the information from both fibers and position it appropriately. The result of the growth of all of the stage fiber is that each simulated cell has the representation of the correct value, plus the horizontal fibers at their eastern edges contain a complete representation of all cell values at this stage of the simulation.  Note that the shaded grey squares are simply filled by generic ``filler tiles'' which carry no specific information.

\begin{figure}[htp]
\begin{center}
\includegraphics[width=3.5in]{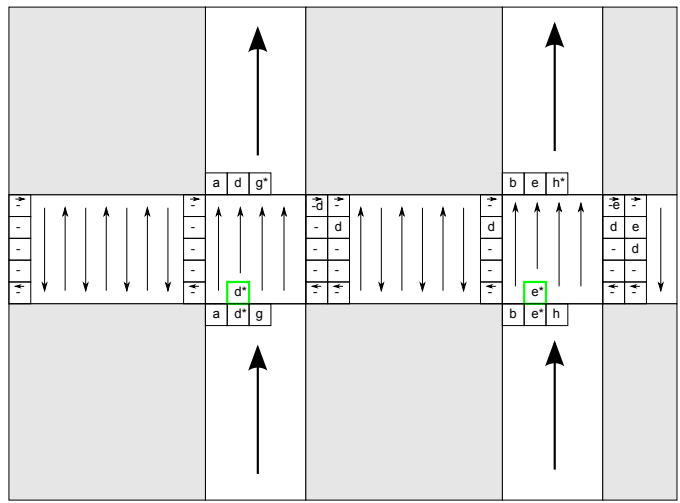}
\caption{Growth of a horizontal fiber, beginning from its initiation and growing through two intersections with vertical fibers.}
\label{fig:TASsimCA-horiz-fiber}
\end{center}
\end{figure}

The entire top row of fiber of a stage is specially designated and grows to a height of $1$ less than the others as it grows from left to right.  The completion of the first stage occurs when the top rightmost square (that representing cell $i$ in Figure~\ref{fig:TASsimCA-stage1done}) has its top rightmost corner completed.  Once this tile is placed, it allows for the attachment of the tile shown in black, which initiates a row of tiles which grow to the left edge of the assembly, and then down the entire left side.  Note that since both the top and left fibers were narrower than the other fibers by a single tile, they are now the same widths, creating a perfectly square assembly for the representation of the initial stage.

\subsection{Growth of subsequent stages}

\begin{figure}[htp]
\begin{center}
\includegraphics[width=3.5in]{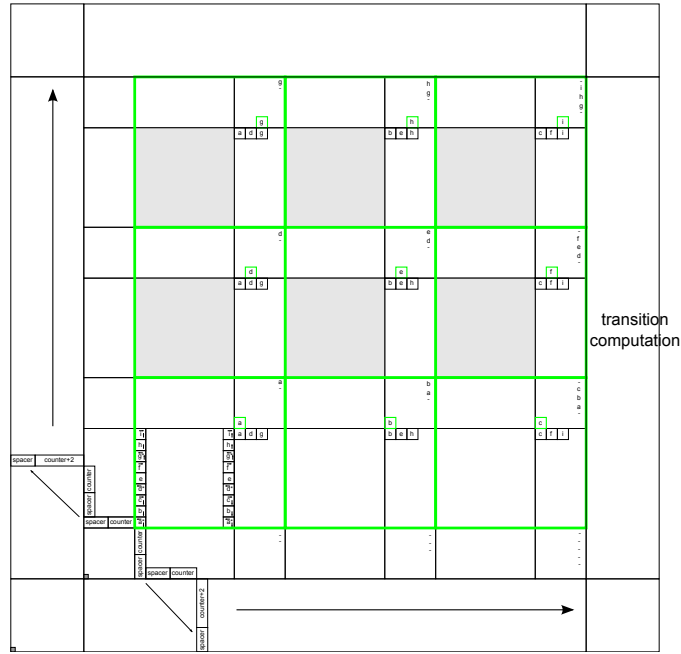}
\caption{Beginning growth of the second stage.}
\label{fig:TASsimCA-stage2}
\end{center}
\end{figure}

To begin growth of the next stage, upon reaching the southwest corner of the assembly representing the first stage, growth is initiated along both the south and west sides which copies the counter and spacer information from the counters there, and then adds $2$ to the maximum counter value (since the dimensions of the cells simulated by this stage will increase by $2$ to add a new perimeter of quiescent cells to handle any growth of the active configuration).  This results in the new boundary fiber for the next stage, and these counters increment each time they reach a square representing the maximum length of the boundary fiber for the previous stage.  Upon reaching that distance, the bottom boundary fiber creates a square which initiates upward growth of a special stage fiber.  While quite similar to the corresponding stage fiber of the previous stage fiber, rather than just extracting the necessary values for the first column of this stage, this fiber embeds the functionality of the ``transition computation gadget.''  This functionality is described in detail in Section~\ref{sec:TASsimCA-comp-gadg}, but essentially it is used to compute the new values of the cells in the current column.  Other than the fact that the first row of vertical stage fibers for each stage use the transition computation gadget rather than just cell value selection as for the first stage, the rest of the information propagation through the fibers is the same.

The vertical stage fiber is able to use the cell values exposed by the horizontal fibers of the last stage to compute the new cell values, with each such horizontal fiber exposing the values of one row of cells.  The leftmost vertical stage fiber places the ``$*$'' marker on the bottom value of each, which represents the rightmost cell of each row.  It then performs the computation of new values and passes them upward by rotating them up and right, with the ``$*$'' remaining on the leftmost.  Note that there is room to accommodate all $n$ values after performing the computation because the computation requires two columns and the width of the fiber is $n+2$.  While doing this, the vertical fiber also passes all of the values from the horizontal fibers of the previous stage through from left to right, while moving the ``$*$'' marker for each up by one position.  This makes the values available for the computation of the next column's values by the next vertical fiber, with the ``$*$'' markers in the correct positions.
When a vertical fiber reaches the topmost row of simulated cells for the current stage, it will insert a new quiescent value row of newly added cells at the top of the grid during this step, and mark them as the top cells of the grid.  The fact that the values of these cells were not included in the computation of cell values for this stage cannot result in incorrect computation due to the fact that the initial configuration $c'_0$ was created with a perimeter of quiescent cells, and growth at every step has ensured that there is always a buffer of quiescent cells which can be assumed during the computation.

The vertical fiber is able to determine how high to grow by detecting the marker from the top right corner of the previous stage.  It is able to determine which information is from the immediately previous stage (in the case of advanced stages where there are stage fibers from multiple previous stages available) because, as a vertical fiber for one stage copies across the values for the fibers of previous stages, if it is the last vertical fiber for its stage (which can easily be determined by the boundary fiber which initiates its growth) it marks the information it is copying across from all other fibers as ``copy-only'', meaning that that information no longer participates in the computation of new cell values.  The vertical fibers within the square representing the computation of values for a given stage copy all information from previous fibers upward and to the right.  As the next stage forms, all such information is copied into and through the neighboring cells.  If it is received from below, it is copied up and to the right, and if it is received from the right it is only copied to the right.

This system allows fibers of each stage to pass along the fibers and values from all previous stages, and only when they are vertical fibers of the initial formation of a stage do they perform computation of new cell values.  Otherwise, fibers are simply nested copies carrying the initial computation of the cell values of their stage, which are passed around through the cells of later stages.  This ensures that the value for each cell at a particular time step is only computed once (which is necessary in the case of the simulation of a nondeterministic CA, to be discussed), even though it is represented (for that cell and time step) in an eventually infinite number of cells (as recursive copies of all previous configurations leading up to each cell).  Thus, each cell of the computation at time step $n$ contains the full configuration history of $\mathcal{A}$ for steps $0$ through $n-1$.

\subsection{Transition computation gadget}\label{sec:TASsimCA-comp-gadg}

We now define the \emph{transition computation gadget}, which is used to execute the transition function which computes the new value of a column of cells given the current values of all cells in the grid.  This consists of a pair of vertical columns which grow up then down along the length of a column of tiles which expose the values of the simulated cells from the previous stage.  These will be contained in the ends of the fibers of the immediately previous stage, with each fiber containing the values of exactly one row, in row order of left to right.  Further, the values of any cells which are on the boundary of the grid will be marked accordingly.  For the first (leftmost) transition computation gadget of each stage, the bottom values of each row will be assumed to implicitly have a ``$*$'' mark, and this mark will be shifted upward by one position for use by each subsequent transition computation gadget.  The purpose of the first, upward growing column is to gather the values of the following cells of the neighborhood of each cell marked with a ``$*$'':  $\{(-1,-1),(0,-1),(1,-1),(-1,0)\}$.  (Recall that the locations marked with a ``$*$'' represent the values for the cells in the current column.)  For clarity, we will now explain how it does this for a single ``double marked'' location (i.e. just for the case of this explanation, one of the locations marked with a ``$*$'' is also given another mark, say ``$+$'') , as the process can easily be extended to simultaneously handle all of the ``$*$'' marked tiles but the explanation is a bit clearer when focusing on one of them.  Computation for the single position is accomplished by the upward growing row ``remembering'' the values of the last cell encountered until it encounters a `$*$', then remembering the last cell, the cell with the `$*$', and the next cell.  It then continues by remembering those three and the last encountered cell until it either encounters another `$*$', at which point it forgets the last group of 3 and starts over, or it encounters the specially ``double marked'' cell.  Once it reaches that cell, following the scheme outlined it will have arrived carrying the necessary information for the four neighbors.  (Note that when tiles record the fact that a cell is on the border of the grid, then the quiescent value can be substituted for the missing neighbor cell or cells.)  The upward growing column completes, then initiates the downward growing column which similarly gathers the value of the remaining four neighbors.  The correct location of the `$*$' markings is crucial for this to occur correctly.  All neighborhood information can be gathered since only a constant amount of information must be contained in any given tile.  See Figure~\ref{fig:TASsimCA-transition-gadget} for two examples of the values for a cell's neighborhood being gathered.  Again, note that the process can be easily extended so that all cells marked with ``$*$'' are computed in the same two rows by just gathering and ``dropping off'' the information at all relevant locations, while still retaining the need for only a bounded number of cell states to be remembered by any single tile, regardless of the size of the grid and thus the number of marked tiles at any time.

Once the downward growing column reaches the height of the cell to transition, the glues adjacent to that location contain the information about the entire neighborhood.  By having one tile type for each possible set of neighborhood values, namely all $2^9$ combinations of $0$'s and $1$'s for the $9$ cells in a neighborhood, the tile set is designed to allow for the placement of exactly one tile type, which correctly represents the value of that cell if it executed its transition function.

\begin{figure}[htp]
\begin{center}
\includegraphics[width=2.5in]{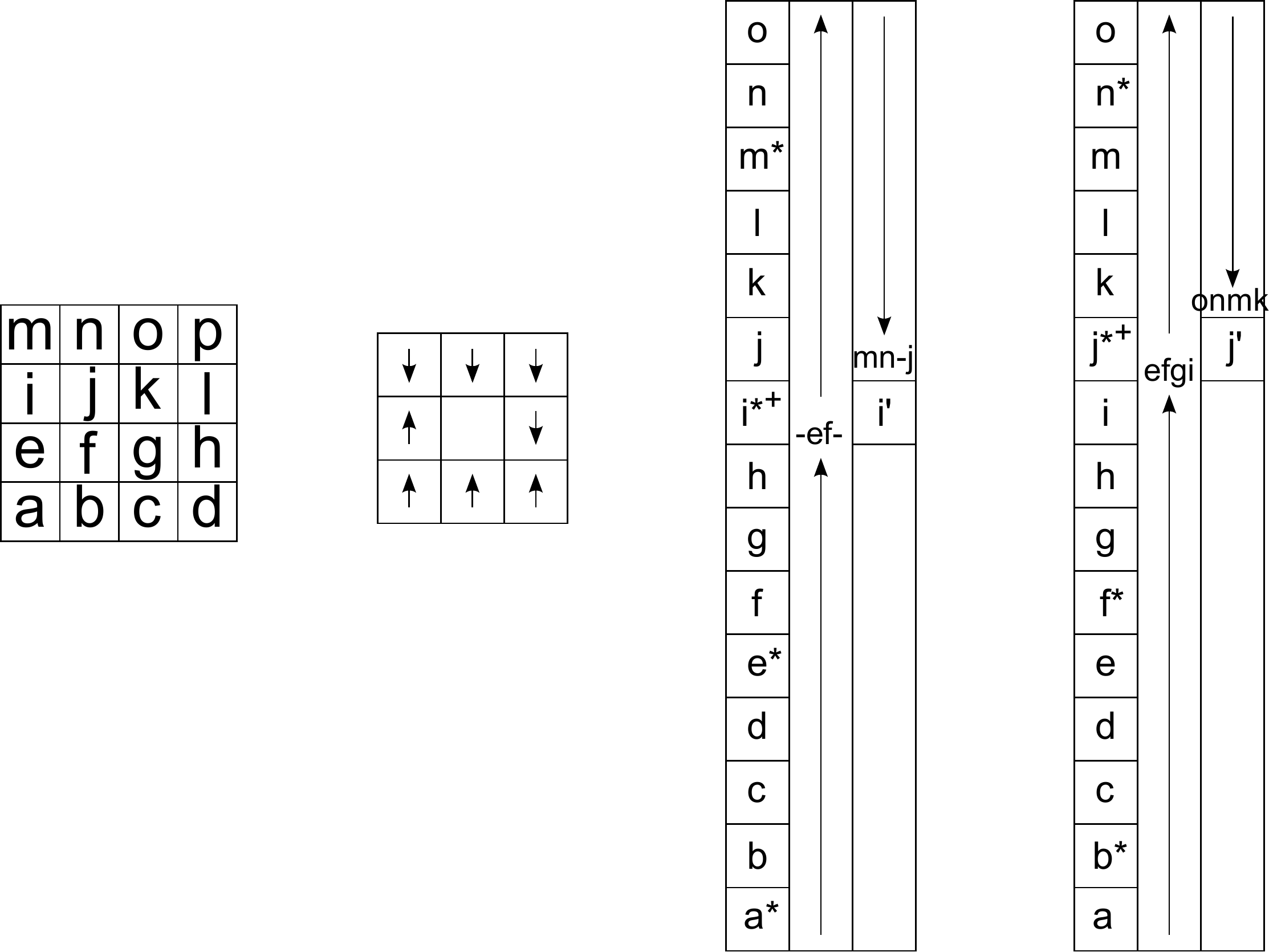}
\caption{The functioning of the transition computation gadget.  (Left) The example grid being considered, (Middle-left)  The neighborhood surrounding a cell and for each neighbor, an arrow indicating whether the upward or downward growing column is used to collect its value. (Middle-right) Gathering the neighborhood values for cell $i$. Note that since $i$'s column is on the left side of the grid, the markings on the cells of that column (previously discussed but not shown here) allow the quiescent value to be used for the cells currently off of the simulated grid. (Right) Gathering neighborhood values for cell $j$.}
\label{fig:TASsimCA-transition-gadget}
\end{center}
\end{figure}

\subsection{Correct simulation}

The scalable representation function $R$ which testifies to the simulation of $\left(\mathcal{A}, c_0\right)$ by $\mathcal{T}$ works as follows.  Given a time step $t \in \mathbb{N}$, it is able to find the scale factor at which the configuration $c \in G^t(c_0)$ is represented by first finding the scale factor of the first stage ($t=0$), which is $n^3 + (n+1)(n+2+\lambda)$ where $n$ is the dimensions of $c'_0$.  Let $w_t = n + (2+\lambda)(t+1)$ be the width of the fiber at step $t$, and we can recursively define the dimensions at step $t$ as $d_t = (n+1)w_t + nd_{t-1}$.  After computing $d_t$, $R_{d_t}$ can simply inspect the given assembly $\alpha$ to determine if the completed square of the necessary dimensions exists.  If so, it can use the computed dimensions to find the locations of the square intersections of stage fibers where the cell values for each simulated cell are marked to determine the states of all cells in the grid being simulated for step $t$.  All other cells of $\mathcal{A}$ must be in the quiescent state.  If not, $R_{d_t}$ is undefined, which ensures that the representation of the CA proceeds in a synchronous manner, with each time step being defined only once all cells states have been computed. For $t$ in $\mathbb{N}$, scalable representation function $R$ and function $f(t) = d_t$, we can see that the conditions of simulation given in Definition~\ref{def:tassim} hold. As an interesting side effect, all simulated cells for time step $t$ contain the entire computational history of $\mathcal{A}$ for time steps $0$ through $t-1$.

}  %

\subsection{Overview of construction to prove Theorem~\ref{thm:TAMsimIU}}

Now that we have defined the above construction for a tile set which can simulate the Game of Life CA given an arbitrary finite initial configuration, we sketch the necessary extensions to provide a tile set which can simulate \emph{any} synchronous nondeterministic CA.

Let $\mathcal{C} = (\mathbb{Z}^2, S, N, \delta)$ be an arbitrary synchronous nondeterministic CA, $d$ be the maximum unit distance of any element of $N$ from the center position (i.e. the distance of a cell's furthest neighbor in its neighborhood), $b$ be the number of bits required to represent $S$, and $c_0$ be the initial configuration of $\mathcal{C}$.  Now, define $M$ as a nondeterministic Turing machine which takes as input the encoding of a synchronous nondeterministic CA (in some standard encoding) and the representation of a grid of cells in the same format they are represented in the previous construction (i.e. as they appear in the seed assembly or along the western edges of the stage fibers of a given stage), with the only difference being that the states are not now restricted to only single bits, but instead may be series of bits (with delimiters between the bits representing the states of different cells), and outputs the new cell state for the one cell marked with ``$*$'' and ``$+$''.  Note that $M$ must be nondeterministic to simulate $\mathcal{C}$, and in order to randomly select from a set of $s$ possible states it simply chooses the bits of a binary number $i$ between $0$ and $s-1$ (the random choice of bits will be performed by the nondeterministic attachment of one of two tiles in each bit position) and then selects the $i$th of the possible states.  For details on how to approach uniform distribution across the selection of possible choices, various gadgets of increasing but bounded space complexity can be used (see \cite{RNSSA}).  Define $r$ to be the longest running time of $M$ when given any single neighborhood $S^{|N|}$ for $\mathcal{C}$, and let $m$ be the maximum amount of tape space used.  Note that both $r$ and $m$ can be easily (if exhaustively) determined by simply running $M$ for each of the $S^{|N|}$ possible neighborhoods.

Now we adjust the previous construction as follows.  To create $c'_0$ from $c_0$, we do as before, but we add an additional $d$ rings of quiescent cells around $c'_0$ to account for the fact that $\mathcal{C}$ may have an arbitrarily large neighborhood and we want to simulate enough quiescent cells around the border of our grid at all times to ensure that we are simulating all non-quiescent cells.  In the seed assembly, in place of the ``spacer'', encode the definition of $\mathcal{C}$ (in the encoding used by $M$) plus $r+m$ spacer tiles to provide enough space for $M$'s tape and running time (the space provided here will ensure that rotations of these values provide the necessary space throughout the assembly).  In place of the transition computation gadget, $M$ will be run.  In order to do this, whenever the boundary fiber reaches a point at which it would have formerly grown a square which initiates the transition computation gadget, instead of growing the square, $M$ is simulated in a standard zig-zag manner.  Additionally, $M$ only computes the transition for a single cell (beginning with the bottom one marked with ``$*$'') and then passes the newly computed cell value along with the rest of the (unchanged) cell values for that column upward.  As before, all needed information is also passed through the simulation of $M$ to the right.  Now, rather than just receiving the new cell values and passing them along, the squares at the intersections of vertical and horizontal stage fibers also execute $M$, whose definition is passed in from the west via the western boundary fiber.  This allows all of the same information flow, but splits up the computations of new cell values in such a way that each simulated cell computes at most a single new cell value, which can be done within the time and space bounds, $r$ and $m$.  In order to provide consistency of scale, a counter is embedded within the running of $M$ so that, in the case of computations which require variable amounts of time relative to each other, the counter ensures that $r+m$ space is always used.  Again, as in the previous construction, once the value for a cell at a particular time step is computed, it is continually passed along into all other representations of the same cell within other larger cells, but never recomputed.

Finally, the representation function $R$ now must be adjusted to take into account the fact that cell states are now binary strings, and also take into account the new scaling factor due to the padding provided to run $M$.  Nonetheless, this construction retains the same properties such as the previous, in terms of completing each stage before beginning the next.

\section*{Acknowledgements}
The authors would like to especially thank Damien Woods and Pierre-Etienne Meunier for valuable discussions, guidance, and suggestions.

\ifabstract
\later{
}
\fi
\iffull
\fi
\bibliographystyle{abbrv} %
\bibliography{tam,ca,ca-USA}

\newpage

\appendix
\magicappendix

\end{document}